\documentclass{jfm}
\usepackage{graphicx}
\usepackage{tikz}
\usetikzlibrary{shapes}
\usepackage{epstopdf, epsfig}
\usepackage{amsmath}
\usepackage{color}
\usepackage{float}
\usepackage{placeins}
\usepackage{subcaption}
\usepackage{natbib}
\usepackage{bm}
\usepackage{booktabs}
\usepackage{caption}
\usepackage{siunitx}
\definecolor{magenta}{rgb}{0.79216,0.12156,0.48236}

\newcommand{\uu}{\mbox{\boldmath $u$} {}}

\newcommand{\ee}{\mbox{\boldmath $e$} {}}
\newcommand{\vv}{\mbox{\boldmath $V$} {}}
\newcommand{\xx}{\mbox{\boldmath $x$} {}}

\newcommand{\ttau}{\mbox{\boldmath $\tau$} {}}
\newcommand{\ppi}{\mbox{\boldmath $\pi$} {}}
\newcommand{\tuu}{\mbox{\boldmath $\tilde{u}$} {}}
\newcommand{\tvv}{\mbox{\boldmath $\tilde{V}$} {}}

\newcommand{\tp}{\mbox{$\tilde{P}$} {}}

\newcommand{\tn}{\mbox{$\tilde{n}$} {}}

\newcommand{\overbar}[1]{\mkern 1.5mu\overline{\mkern-1.5mu#1\mkern-1.5mu}\mkern 1.5mu}

\newcommand{\blackldline}{\raisebox{2pt}{\tikz{\draw[-,black,dash pattern=on 3pt off 6pt,line width = 0.9pt](0,0) -- (5mm,0);}}}

\newcommand{\redddline}{\raisebox{2pt}{\tikz{\draw[-,red,dotted,line width = 0.9pt](0,0) -- (5mm,0);}}}
\newcommand{\redline}{\raisebox{2pt}{\tikz{\draw[-,red,solid,line width = 1.5pt](0,0) -- (5mm,0);}}}

\newcommand{\blueddline}{\raisebox{2pt}{\tikz{\draw[-,blue,dotted,line width = 1.5pt](0,0) -- (5mm,0);}}}

\newcommand{\greendline}{\raisebox{2pt}{\tikz{\draw[-,black!60!green,dashed,line width = 1.5pt](0,0) -- (5mm,0);}}}

\newcommand{\reddashdline}{\raisebox{2pt}{\tikz{\draw[-,red,dash pattern={on 5pt off 1pt on 2pt off 1pt on 5pt},line width = 1.5pt](0,0) -- (5mm,0);}}}

\newcommand{\magdline}{\raisebox{2pt}{\tikz{\draw[-,magenta,dash pattern={on 5pt off 1pt on 2pt off 1pt on 5pt},line width = 1.5pt](0,0) -- (5mm,0);}}}

\newcommand{\rtri}{\raisebox{0.5pt}{\tikz{\node[draw,red,scale=0.4,regular polygon, regular polygon sides=3,fill=none](){};}}}
\newcommand{\mtri}{\textcolor{magenta}{\rhd}}

\newcommand{\bsq}{\raisebox{0.5pt}{\tikz{\node[draw,blue,scale=0.6,regular polygon, regular polygon sides=4,fill=none](){};}}}

\newcommand{\gcircle}{\raisebox{0.5pt}{\tikz{\node[draw,black!60!green,scale=0.4,circle,line width = 1.0pt,fill=none](){};}}}

\shorttitle{Crossing trajectory effects in LES}
\shortauthor{A. K. Aiyer and C. Meneveau}

\title{Resolved and subgrid-scale crossing trajectory effects in Eulerian large eddy simulations of  {size-dependent} droplet transport}
\author{A. K. Aiyer\aff{1}
  \corresp{\email{aaiyer@princeton.edu}} \and C. Meneveau\aff{2}
 }

\affiliation{\aff{1}Department of Mechanical and Aerospace Engineering, Princeton University,
Princeton, NJ 08544, USA
\aff{2}Department of Mechanical Engineering, Johns Hopkins University, Baltimore, MD 21218, USA}

\begin{document}

\maketitle

\begin{abstract}

We study the dispersion characteristics of slightly buoyant droplets in a turbulent jet using large eddy simulations (LES). The droplet number density fields are represented using an Eulerian approach with the dispersed phase modelled using the Fast-Eulerian method \citep{Ferry2001} that includes the 
droplet rise velocity. Radial concentration profiles and turbulent concentration fluxes for  droplets of different sizes are analyzed to quantify the ``trajectory crossing effect'', when relative motions between particles and turbulent eddies tend to reduce turbulent diffusion. For finer LES grid resolutions, the model captures the differential, size based dispersion characteristics of the droplets with the transverse dispersion of the larger droplet sizes suppressed, since trajectory crossing effects are explicitly resolved in LES.  We examine a similarity solution model for the size dependent  radial concentration profiles based on a modified Schmidt number derived from the theory of turbulent diffusion of particles in the atmosphere proposed by \citet{Csanady1963}.  The results are validated with the high resolution LES data and show good agreement. Then, the size dependent Schmidt number model is reformulated as a model for unresolved subgrid-scale trajectory crossing effects and used to calculate the subgrid concentration flux in a coarse LES of a turbulent jet, with slightly buoyant droplets injected at the centerline in the self-similar region of the jet. The results are compared to a simulation with higher grid resolution and a coarse simulation with a constant Schmidt number SGS model. We find that the subgrid model enhances the prediction accuracy of the concentration profiles and turbulent concentration flux for the coarse LES. 
  
\end{abstract}


\section{Introduction}
Understanding the behaviour of dispersed particles in turbulent flows 
is relevant in many contexts. In modelling oil spills, knowledge of the oil droplet dispersion characteristics is important in  determining their transport \citep{North2015,Yang2016}. The oil slicks are broken down into polydisperse droplet distributions that are transported by waves and turbulence. The smaller droplets are more horizontally dispersed by the turbulence while the larger ones rise towards the surface.  In studying disease transmission, the dispersion characteristics and size distribution of particles ejected from a sneeze or cough affects their residence time and transport properties in the air, information necessary to formulate  social distancing and masking rules 
\citep{Bourouiba2020,mittal_2020}. There have been considerable efforts in studying dispersion of bubbles and heavy particles in turbulence \citep{corrisin_lumley_1971,Csanady1963,snyder_lumley_1971,Reeks1977OnFluid,Wells1983,Wang_stock_1992,stout_nldrag,Kennedy1998,Mazzitelli2004}. Polydispersity adds an additional layer
of complexity in characterizing particle dispersion due to the interplay of inertial effects (lift forces on different sized particles) and buoyancy effects that can affect particles of different sizes quite differently. The particle-turbulence interactions result in interesting phenomena such as preferential concentration \citep{Squires1991,Eaton1994},  particle clustering \citep{Obligado2014,Falkinhoff2020}  (either strain or vorticity dominated) and modified turbulent diffusion due to particle trajectory crossing  \citep{Yudine1959,Csanady1963,Wells1983}. This effect refers to the relative motion between particles and turbulent eddies in the continuous phase. 

\citet{Yudine1959} and \citet{Csanady1963} studied the case of finite particle rise velocity and zero particle inertia. 
The velocity correlation seen by the heavy particle $\langle u^p_i(0) u^p_j(\tau)\rangle$, can be approximated by the Eulerian spatial velocity correlation of the flow $\langle u_i(\bm{x_0},t_0) u_j(\bm{x_0} + \bm{W_{r}}\tau,t_0)\rangle$ {for small particle correlation time $\tau$},  where $u_i(\bm{x},t)$ is the ith component of the fluid Eulerian velocity. The effect of the rise velocity $W_r$ on the velocity correlation was termed as the "crossing-trajectories effect". \citet{Csanady1963} found that the effect of increasing finite rise/free fall velocity was to reduce both the transverse and longitudinal ({with respect to} the free fall direction) dispersion coefficients. 
\citet{Wang_stock_1992} presented a comprehensive analysis for the dispersion coefficients of heavy particles in isotropic turbulence by separating the effects of inertia and particle rise velocity. They defined a non-dimensional inertia parameter (based on the particle Stokes number) and a rise parameter (based on the rise/fall velocity) to characterize the dispersion of heavy particles. Particles dispersed faster than the fluid elements if the inertia parameter controlled the dispersion (large Stokes number), and slower than the fluid elements if the drift parameter governed the dispersion. The finite free fall (or rise) velocity of particles causes them to migrate out of eddies before they decay \citep{Yudine1959}. This results in particles losing their velocity correlation more rapidly than fluid elements. The particle dispersion is thereby reduced as the particle correlation times are directly related to their dispersion coefficient \citep{Taylor1920DiffusionMovements}.
The following relations were obtained for the normalized particle dispersion coefficients by \citet{Wells1983}: 
 \begin{subeqnarray}
       \frac{D_{p,L}}{D_{f,L}} = \left(1 + \frac{C W_{r}^2}{w^{'2}}\right)^{-1/2}, \\
            \frac{D_{p,T}}{D_{f,T}} = \left(1 + \frac{4C W_{r}^2}{w^{'2}}\right)^{-1/2},
 \end{subeqnarray}\label{eqn:dispersion_coeff}
where $D_{p,L}$, $D_{f,L}$, $D_{p,T}$, $D_{f,T}$ are the longitudinal and transverse particle and fluid dispersion coefficients, respectively,  $W_{r}$ is the particle rise velocity (in the vertical direction), $C = w' T_L/L_E $ relates the Lagrangian timescale $T_L$ to the Eulerian length scale $L_E$   and $w'$ is the axial turbulent fluctuation velocity (in the direction of the rise velocity).  Note that the crossing  trajectory effect referred to here is the one defined by \citet{Yudine1959} that accounts for the effect of a finite drift velocity on the velocity fluctuations to be distinguished from the different  effect of multivalued particle velocity at a given space-time instant \citep{Laurent2012ANumbers}.

A number of 
experimental studies have focused on oil droplets in water, i.e. cases where particles are only slightly buoyant. For instance, \citet{Friedman2002} have shown that the rise velocities and dispersion characteristics of  slightly buoyant oil-in-water droplets differ significantly from those of bubbles and heavy particles. \citet{Gopalan2008a} studied the dispersion characteristics of oil droplets in water experimentally in isotropic turbulence.  They calculated the turbulent diffusion coefficient by integrating the ensemble-averaged Lagrangian velocity autocovariance and found a dependence of the diffusion coefficient on the ratio of the turbulence intensity and the droplet rise velocity. 

Besides experiments, recent years have seen a significant number of computer simulation based studies of particle laden turbulent flows. Numerical approaches for {the particle phase} can be broadly classified into either Lagrangian models, where individual droplets or droplet clusters are tracked in Lagrangian fashion, or Eulerian models where the distribution of particles is treated as a continuous density field \citep{Balachandar2010,Fox2011,Subramaniam2013}. 
Direct numerical simulations (DNS) {of the continuous phase coupled with a Lagrangian approach  for the} particles provides the highest fidelity,  but have a prohibitively high computational cost due to the large number of particles that need to be tracked and resolved, in addition to the high cost of having to resolve the continuous phase down to the Kolmogorov scale. Large eddy simulations (LES) on the other hand capture the large- and intermediate-scale turbulent motions (based on the grid resolution), and only require modelling of the unresolved subgrid-scale turbulence effects. While the cost of LES is higher than Reynolds-averaged Navier–Stokes (RANS) simulations, LES provides the ability to resolve unsteady spatially fluctuating phenomena at least down to scales of the order of the grid scale.  
The conventional wisdom is that
turbulent transport is dominated by the largest eddies in the flow and therefore, LES neglecting subgrid-scale effects other than the subgrid-scale eddy diffusivity should provide robust and accurate predictions in general. However, 
the direct effect of the subgrid scale velocity  
has already been shown to be important for particle clustering, preferential concentration 
and particle statistics, particularly for small particle time-scale (or Stokes number) when the grid resolution of LES is coarse 
\citep{Armenio1999,Marchioli2017}. In order to account for the SGS fluid field on the particle evolution, numerous models have been developed 
such as based on approximate-deconvolution methods \citep{Stolz2001,Shotorban2005} or 
Lagrangian stochastic models \citep{Mazzitelli2004,Minier2004,Oesterle2004,Shotorban2006,Johnson2018,Knorps20212121}. These models aim to 
describe the unresolved portions of the fluid velocity as seen by the particle. These models include various terms in their parameterization to capture the effect of crossing trajectories, 
particle inertia, flow anisotropy etc. \citep{Berrouk2007,Michaek2013}. The models are 
developed for cases where the particle phase is modelled using a Lagrangian approach with the 
velocity field modelled using an Eulerian model.

There have been efforts in modelling the crossing trajectory  effect for particles in homogeneous isotropic turbulence and shear flows in the presence of an external force field.
\citet{Simonin1993} developed a closure equation for the fluid/particle moments required for dispersed phase modelling in the framework of a  two-fluid model. {\citet{Pozorski1998} modeled the crossing trajectory effect by introducing  a Langevin model for the mean drift velocity.} Moment based methods have also been used by \citet{Salehi2019} to study polydisperse inertial particles. \citet{Oesterle2009} provided an alternative theoretical analysis of the time correlation of the fluid velocity fluctuations  using Langevin-type stochastic models to predict the effects of crossing trajectories in shear flows.
 Using DNS with Lagrangian modeling of heavy particles, \citet{Fede2006} have quantified the effects of the particle trajectory crossing effect on the structure of subgrid-scale turbulence and its time scales.

For applications with relatively low volume fractions and low particle Stokes number, Eulerian approaches for the particle field can be advantageous as they are not limited by the number of particles \citep{Fox2008NumericalMethod,Laurent2012ANumbers,Masi2014,Vie2016}. {\citet{Pandya2002} and \citet{Zaichik2009} formulated a two-fluid LES approach for particle laden turbulent flows. The approach was based on a kinetic equation for the filtered probability density of the particle velocity and was valid for larger Stokes numbers}. In the Equilibrium-Eulerian approach \citep{Ferry2001,Shotorban2007,Shotorban2009} the particle velocities are modelled as a function of the particle timescale to include effects of buoyancy, added mass and subgrid-scale effects. The particles are represented by a continuous concentration field and a separate transport equation is solved for each particle size  \citep{Yang2016,Aiyer2019AEvolution}. LES of polydisperse droplets/bubbles with the equilibrium-Eulerian approach have been used to study bubble driven plumes by \citet{Yang2016} and have been coupled with population balance equations to model droplet breakup by  \citet{Aiyer2019AEvolution,Aiyer2020CoupledJet}.  Due to its advantages for LES of low volume fraction oil-water multiphase flows {where the particle Stokes number is small}, in this paper we focus on   the equilibrium-Eulerian model. Specifically, we aim to explore the degree in which the approach is able to capture effects of particle trajectory crossing and whether for LES performed at very coarse resolutions, additional required subgrid modeling can be developed.

The paper begins in section \S \ref{sec:Eul_sim} describing the Eulerian-Eulerian LES approach.  In section \S \ref{sec:Results}, we present results from a well resolved LES of a turbulent jet with droplets injected at the source of the jet. Results motivate the development of a 
modified-Schmidt number subgrid model to account for crossing trajectory effects. Implementation and validation of the proposed modified subgrid turbulent diffusion model are presented in section \S \ref{sec:subgrid}. Conclusions are presented in section \S \ref{sec:Concl}

\section{Equilibrium-Eulerian large eddy simulation}\label{sec:Eul_sim}
\subsection{Governing equations and numerical methods}

Let $\xx = (x,y,z)$ with $x$ and $y$ the horizontal coordinates and $z$ the vertical direction, and let $\uu = (u,v,w)$ be the corresponding velocity components. Also, $n_i$ is the  
number density of the droplet of size $d_i$. The jet and surrounding fluid are governed by the three--dimensional incompressible filtered Navier--Stokes equations with a Boussinesq approximation for buoyancy effects, while the droplet concentration fields are governed by Eulerian transport equations at each scale:
\begin{equation}\label{eqn:div}
\nabla \cdot \tuu =0,
\end{equation}
\begin{align}\label{eqn:Navier_stokes}
\frac{\partial \tuu}{\partial t} + \tuu \cdot\nabla\tuu =& -\frac{1}{\rho_c}\nabla\tp - \nabla\cdot \ttau^d  +\tilde{F}\,\ee_3\\\nonumber &+\left(1-\frac{\rho_d}{\rho_c}\right)\sum_i(V_{d,i}\tn_i)\,g\,\ee_3.
\end{align}
\begin{equation}\label{eqn:conc}
\frac{\partial \tn_i}{\partial t} + \nabla \cdot (\tvv_i\tn_i) + \nabla \cdot \ppi_{i} = \widetilde{q}_{i}, \,\,\,\, i=1,2.. N.
\end{equation}
A tilde denotes a variable resolved on the LES grid, $\tuu$ is the filtered fluid velocity, $\rho_d$ is the density of the 
droplet, $\rho_c$ is the carrier fluid density, $V_{d,i}=\pi d_i^3/6$ is the volume of a spherical droplet of diameter $d_i$, $\ttau = (\widetilde{\uu\uu} - \tuu\tuu)$ is the subgrid-scale stress tensor (superscript ``d'' denotes its deviatoric part), $\tn_i$ is the resolved 
number density of the droplet of size $d_i$, $\widetilde{F}$ is a locally acting upward body force to simulate the jet momentum injection, and $\ee_3$ is the unit vector in the vertical direction
The droplet  Eulerian description has been used previously to study mono-disperse plumes \citep{Yang2014b,Yang2015,Chen2016,Yang2016} and polydisperse oil plumes \citep{Aiyer2019AEvolution,Aiyer2020CoupledJet}.
The filtered version of the transport equation for the number density $\widetilde{n}_i(\xx,t;d_i)$ is given by Eq. (\ref{eqn:conc}). The term $\ppi_{i} = (\widetilde{\vv_in_i} - \tvv_i\tn_i)$ is the subgrid-scale concentration flux of oil droplets of size $d_i$ (no summation over \textit{i} implied here) and $\widetilde{q}_i$ denotes the injection rate of droplets of diameter $d_i$.
In order to capture a range of sizes the number density distribution function is discretized into $N$ linearly distributed bins and we solve $N$ separate transport equations for the number densities $\widetilde{n}_i(\xx,t;d_i)$  with $i=1,2,...,N$.

Closure for the SGS stress tensor $\ttau^d$ is obtained from the Lilly-Smagorinsky eddy viscosity model with a Smagorinsky coefficient $c_s$ determined dynamically during the simulation using the Lagrangian averaging scale-dependent dynamic (LASD) SGS model \citep{Bou-Zeid}. {In the current simulations the viscous stress contribution has been neglected. The eddy viscosity at the centreline of the jet for the configurations considered here is about $~20$ times the molecular viscosity. Towards the edge of the jet at $r/r_{1/2} = 2$, the value decays to 10 times the molecular viscosity. The viscous stress would be important in the near nozzle region of the jet, a region not included explicitly in the present LES.} The SGS scalar flux $\ppi_{i}$ is modelled using an eddy-diffusion SGS model. 
We follow the approach of \citet{Yang2016} and prescribe a constant turbulent Schmidt/Prandtl number, $Pr_{sgs} = Sc_{sgs} =0.4$. The SGS flux is thus parameterized as $\pi_{n,i} = -(\nu_{sgs}/Sc_{sgs})\nabla\tn_{i}$. 
 With the evolution of  oil droplet concentrations being simulated, their effects on the fluid velocity field are modelled and implemented in  $(\ref{eqn:Navier_stokes})$ as a  buoyancy force term (the last term on the right-hand side of the equation) using the Boussinesq approximation. A basic assumption for treating the oil droplets as a Boussinesq active  scalar field being dispersed by the fluid motion is that the volume and mass fractions of the oil droplets are small within a computational grid cell.
The droplet transport velocity $\tvv_i$ is calculated by an expansion in the droplet time  scale $\tau_{d,i}=(\rho_d+\rho_c/2)d_i^2/(18\mu_f)$ \citep{Ferry2001}. The expansion is valid when $\tau_{d,i}$ is much smaller than the resolved fluid time scales, 
which requires us to have a grid Stokes number $St_{\Delta,i} = \tau_{d,i}/\tau_{\Delta} \ll 1$, where $\tau_{\Delta}$ is the turbulent eddy turnover time at scale $\Delta$.
The transport velocity of droplets of size $d_i$, $\tvv_i$,  is given by \citep{Ferry2001}
\begin{equation}\label{eqn:rise_vel}
\tvv_i = \tuu + W_{r,d_i}\ee_3 + (R-1)\tau_{d,i}\left(\frac{D\tuu}{D t} + \nabla \cdot \ttau\right) ,
\end{equation}
where $\ee_3$ is the unit vector in the 
vertical direction, and $R = 3\rho_c/(2\rho_{d}+\rho_c)$ is the acceleration parameter. {The droplet rise velocity is calculated as a balance between the drag and buoyancy force acting on a droplet.
\begin{equation}
W_{r,d_i} =
    \begin{cases}
      W_{r,S}, & \text{if}\ Re_i < 0.2, \\
      W_{r,S}(1+0.15Re_i^{0.687})^{-1}, & 0.2 <Re_i< 750,
    \end{cases}
  \end{equation}
  where
  \begin{equation}
  W_{r,S} = \frac{(\rho_c-\rho_d)g\ d_i^2}{18\mu_f},
  \end{equation}
  and $Re_i = \rho_c W_{r,d_i} d_i/\mu_c$ is the droplet rise velocity Reynolds number and $g=9.81\ \mathrm{m}/\mathrm{s}$ is the gravitational acceleration.
}
The effects of drag, buoyancy, added mass and the divergence of the subgrid stress tensor have been included in the droplet velocity expansion. A more detailed discussion of the droplet rise velocity in Eq. $(\ref{eqn:rise_vel})$ can be found in \citet{Yang2016}. 
{While in general 
an additional term proportional to the difference of particle and local fluid accelerations 
should be included \citep{climent1999large}, 
\citet{bec2006} showed that for Stokes number, $St < 0.4$ the difference of the rate of variation of particle momentum and rate of variation of the fluid momentum at the particle position can be neglected.}

The equations ($\ref{eqn:div}$) and ($\ref{eqn:Navier_stokes}$) are discretized using a pseudo-spectral method on a collocated grid in the horizontal directions and a centered finite difference scheme on a staggered grid in the vertical direction \citep{Albertson1999}. Periodic boundary conditions are applied in the horizontal directions for the velocity and pressure field.
The transport equations for the droplet number densities, Eq. ($\ref{eqn:conc}$),
are discretized as in \citet{Chamecki2008}, by a finite-volume algorithm with a bounded third-order upwind scheme for the advection term.  A fractional-step method with a second-order Adams--Bashforth scheme is applied for the time integration, combined with a standard projection method to enforce the incompressibility constraint. The same methods were used in \citet{Yang2016,Aiyer2019AEvolution} where more detailed descriptions can be found.

\subsection{Simulation Setup}

\begin{figure}
    \centering
    
     \subfloat{\includegraphics[width = 0.48\columnwidth]{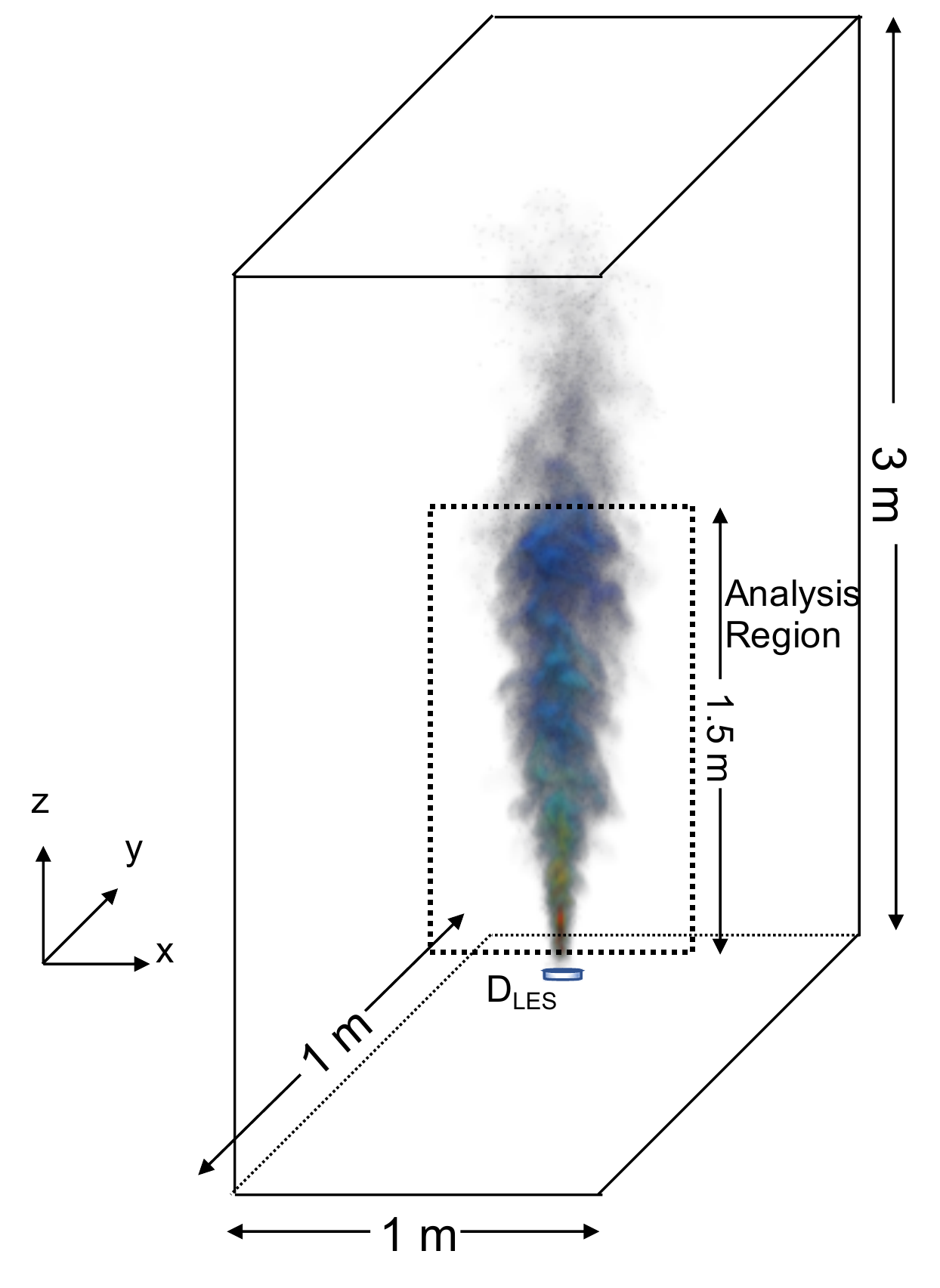}}
     \hfill
    \subfloat{\includegraphics[width = 0.48\columnwidth]{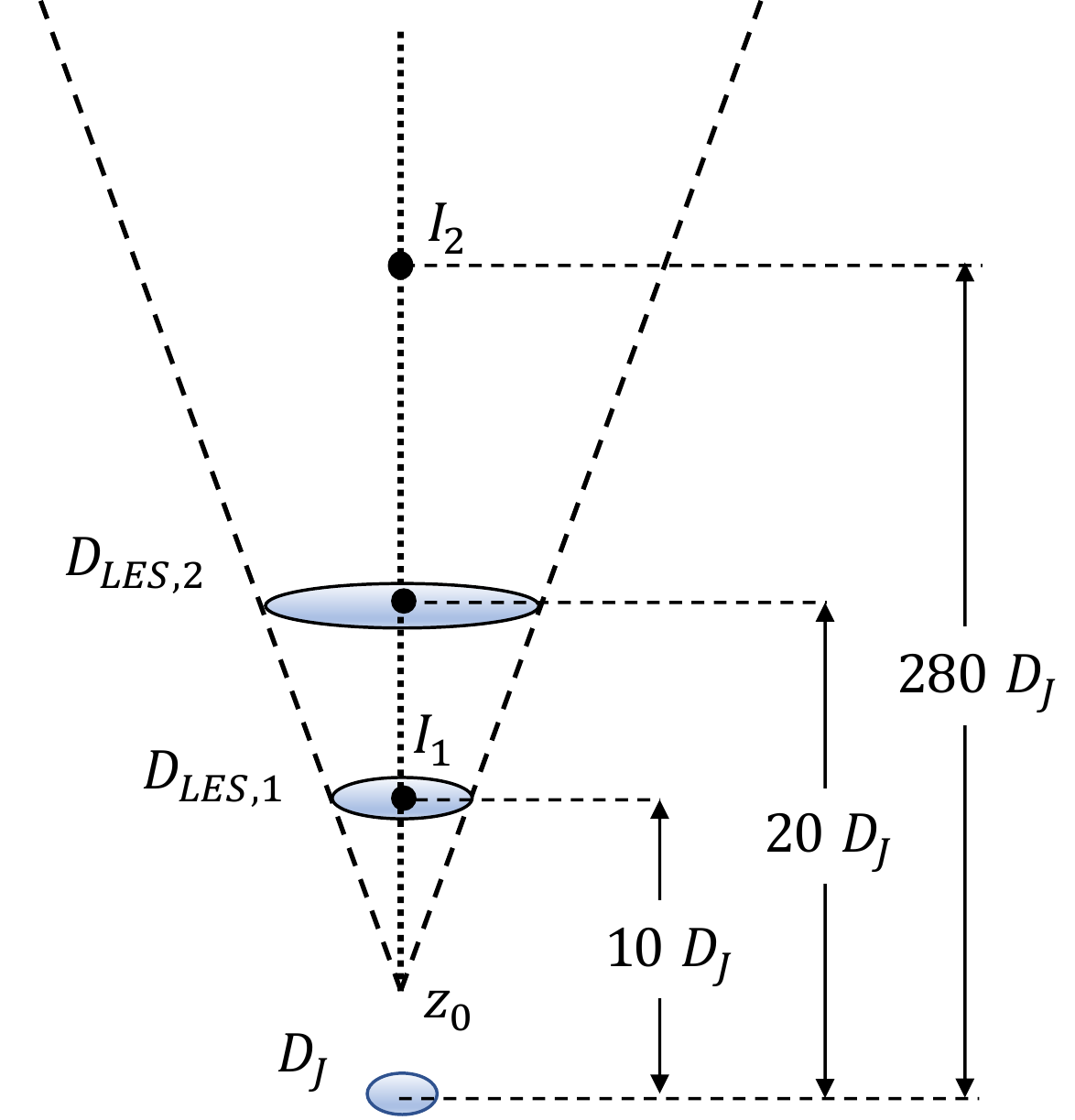}}

  \caption{ Sketch of the simulation setup. Volume rendering of the instantaneous $50\;\mu\mbox{m}$ diameter droplet concentration with the $1000\;\mu\mbox{m}$ droplets visualized as dots placed randomly with density proportional to its concentration field. (b) Sketch depicting the LES nozzles ($D_{LES,1}$, $D_{LES,2}$), the virtual (true) nozzle ($D_v$) and the droplet injection locations ($I_1$, $I_2$)
}
    \label{fig:sketch_sim}
\end{figure}
{
\begin{table}
\centering

\begin{tabular}{ c  c  c  c c} 
 $\rho_d$ (kg/m$^3$) & $\nu_d$ (m$^2$/s) & $\sigma$ (N/m) & $\rho_c$ (kg/m$^3$) & $Re =  {U_JD_{LES}}/{\nu_c}$ \\ [0.5ex] 
 \midrule
   $864$ & $1.02\times 10^{-5}$ & $1.9 \times 10^{-2}$ & $1018.3$ & 12000\\ 
\end{tabular}
\caption{Simulation Parameters.}
\label{tab:oil_prop}
\end{table}
}

The injected jet is modelled in the LES using a locally applied vertically upward pointing body force following the procedure outlined in \citet{Aiyer2019AEvolution,AiyerThesis}. We use this approach since at the LES resolution used in the current applications it is not possible to resolve the small-scale features of the injection nozzle. A sketch for the simulation setup is shown in Figure \ref{fig:sketch_sim}. The body force is applied downstream of the true nozzle ($D_J$ in Figure \ref{fig:sketch_sim}) at a location $z = 10 D_J$, where the actual jet being modeled would have grown for the LES grid resolution to be sufficient to resolve its large-scale features. The strength of the body force is specified so as to obtain an injection velocity of $2\; m/s$. Random fluctuations are added to the horizontal components of the momentum equation to induce transition to turbulence. The fluctuations have an amplitude with a root-mean-square value equal to $0.1$ \% the magnitude of the forcing $\tilde{F}$ and are applied only during an initial period of $0.5\ s$ at the forcing source. The forcing is only applied over a finite volume and 
smoothed using a super-Gaussian kernel. {A summary of important simulation parameters is provided in Table \ref{tab:oil_prop}.}
The first set of simulations  use a relatively fine grid with $N_x \times N_y\times N_z = 288 \times 288 \times 384$ points for spatial discretization, and a timestep
$\Delta t = 3 \times 10^{-4}$ s for time integration. The resolution in the horizontal directions, $\Delta x = \Delta y = 3.47$ mm is set to 
ensure that at the location where the LES begins to resolve the jet we have at least 3 points across the jet. In the vertical direction we use a grid spacing of $\Delta z = 6.5$ mm enabling us to capture a domain height 2.5 times the horizontal domain size. 
This configuration has been validated with DNS data and shown to be robust in producing realistic turbulent round  jet statistics \citep{Aiyer2020CoupledJet}.

The droplets are injected at the source location $I_1$ (see Fig. \ref{fig:sketch_sim}) and the droplet number density fields are initialized to zero in the rest of the domain. {The initial droplet size distribution is uniform and the initial droplet velocity is the rise velocity defined in Eq. (\ref{eqn:rise_vel}).} In order to avoid additional transient effects, the concentration equations are solved only after a time at which the jet in the velocity field has reached near the top boundary to allow the flow to be established. The number density transport contains a source term, $\widetilde{q}_i$ on the RHS of Eq. (\ref{eqn:conc}) 
that represents injection of droplets of a particular size. The droplet size range is discretized into $N = 10$ sizes spanning a range of $d_1 = 50\ \mu m$ to $d_{10} = 1\ mm$. Droplets of different sizes are injected with a constant source flux {$q_i = \tilde{q}_iV_{d,i} = 0.1\; L/min$ }corresponding to a total volume flux, $Q_0 = 1\; L/min$
The source is centered at ($x_c$, $y_c$) = (0.5 m, 0.5 m) and distributed over two grid points in the z direction with weights $0.7$  and $0.3$ at $z_c$ and $z_{c}+\Delta z$ respectively and over three grid points in the horizontal directions with weights $0.292$ at $(x_c,y_c)$ and $0.177$ at $(x_{c}\pm \Delta x,y_{c}\pm \Delta y)$. {The initial droplet size distribution does not effect the properties of the jet due to the small volume fraction of the dispersed phase \citep{Aiyer2020CoupledJet}}.

\section{LES of droplet {transport}: size-dependent turbulent diffusion}\label{sec:Results}

We first examine instantaneous velocity and concentration contours on the mid y-plane as a function of $x$ and $z$ in Figure \ref{fig:inst_profiles}. The left panel depicts the instantaneous vertical velocity, while the right panel shows the concentration of the $d=50\ \mu m$ droplet and the contour lines of the $d=1000\ \mu m$-sized droplets. Visually we can already see that the smallest droplets' spatial distribution has a wider horizontal extent as we move further from the injection location as compared to the width of the distribution of larger droplets.

\subsection{Mean velocity and concentration profiles}

The statistics of the velocity and concentration fields are presented using a cylindrical coordinate system with $z$ being the axial coordinate, and supplement the time averaging with additional averaging over the angular $\theta$ direction around the jet axis. The LES data on the Cartesian grid are interpolated using bilinear interpolation onto a much finer  $\theta$ grid in the horizontal directions to generate smoother profiles after averaging over the $\theta$ direction. 
\begin{figure*}
 \hspace{-1cm}
\centering
        \subfloat[Velocity\label{fig:inst_vel}]{\includegraphics[width=0.54\columnwidth,trim = -10 -2 -2 -2]{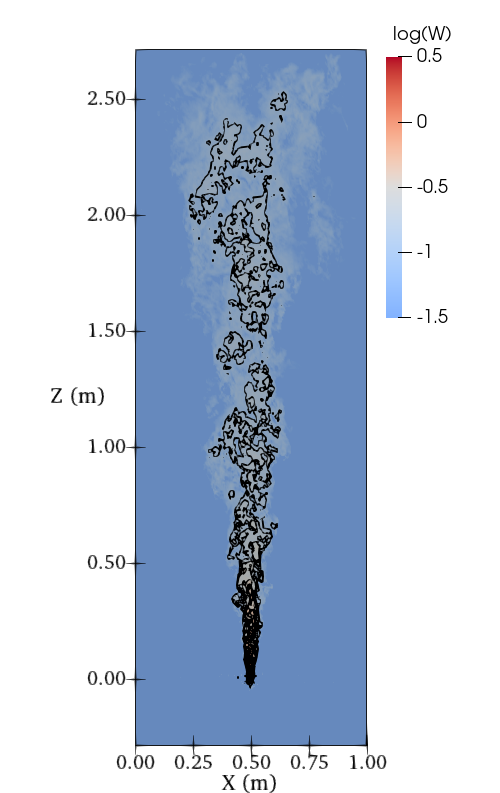}}
    \hfill
\subfloat[Concentration\label{fig:inst_conc}]{\includegraphics[width=0.52
\columnwidth,trim = -2 -2 -2 -2]{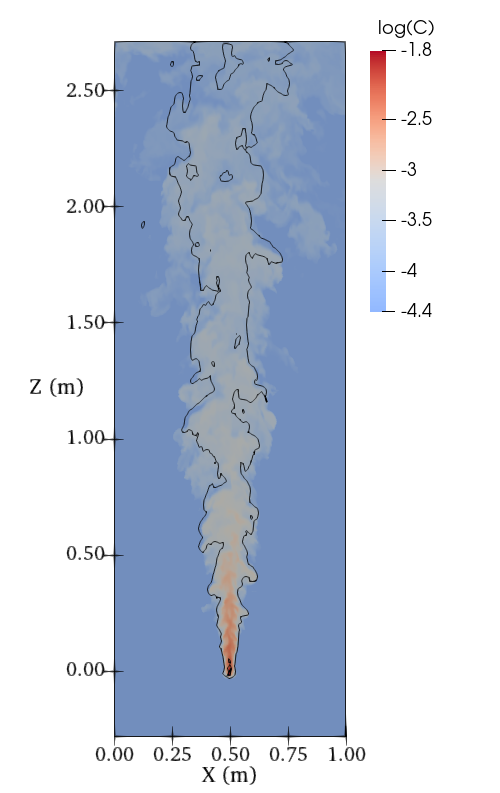}}
\caption{\protect\subref{fig:inst_vel}) Instantaneous snapshot of the velocity at the midplane of the jet in logarithmic scale , \protect\subref{fig:inst_conc}) Instantaneous concentration contours for $d= 50\ \mu m$ (color) and $d = 1\ mm$ (lines, {contour level $\tilde{c}=10^{-4}$}) at the midplane of the jet}
\label{fig:inst_profiles}
\end{figure*}

\begin{figure*}
 \hspace{-1cm}
\centering
        \subfloat[Velocity\label{fig:ss_vel}]{\includegraphics[width=0.48\columnwidth]{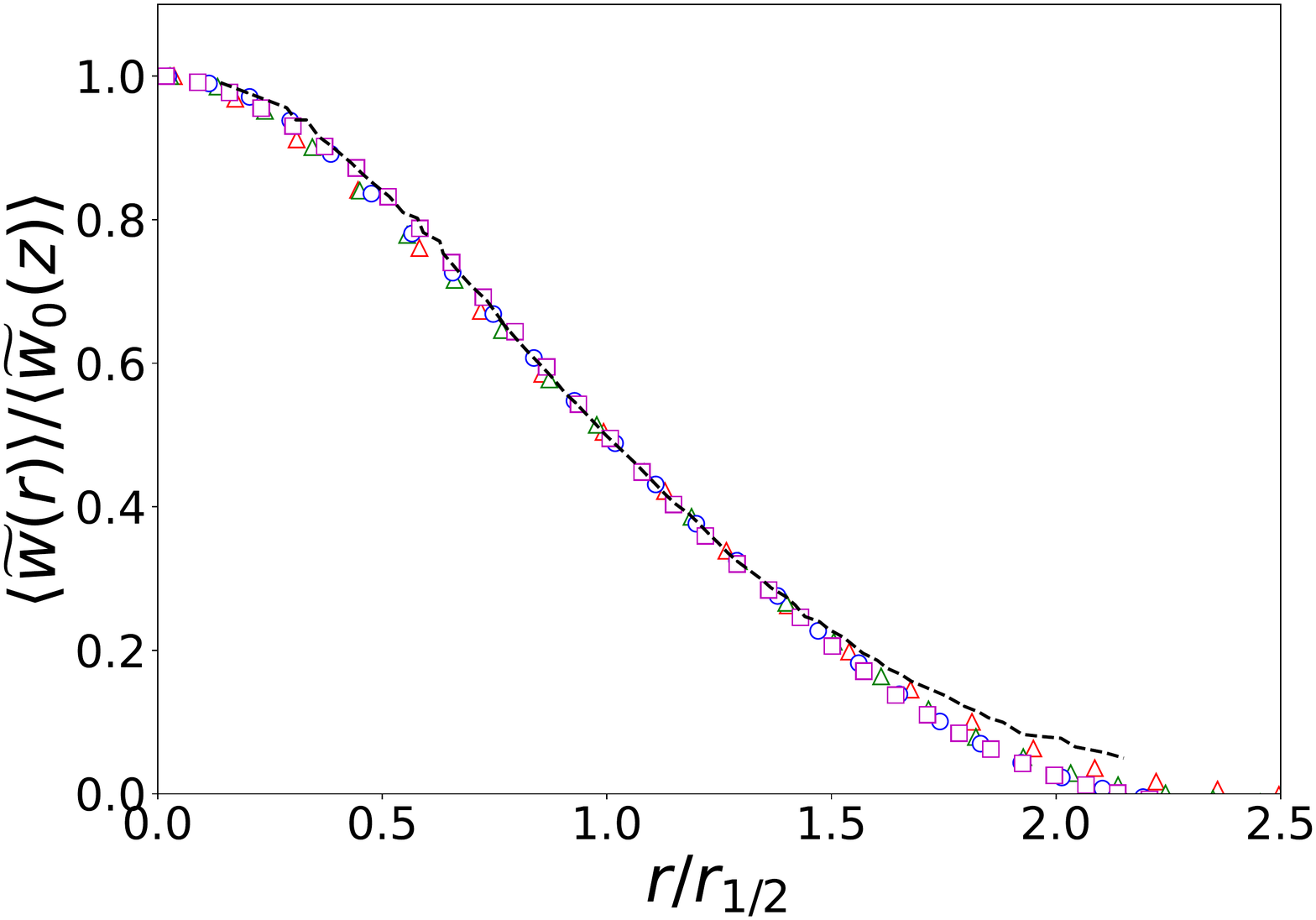}}
    \hfill
\subfloat[Concentration\label{fig:ss_conc}]{\includegraphics[width=0.48
\columnwidth]{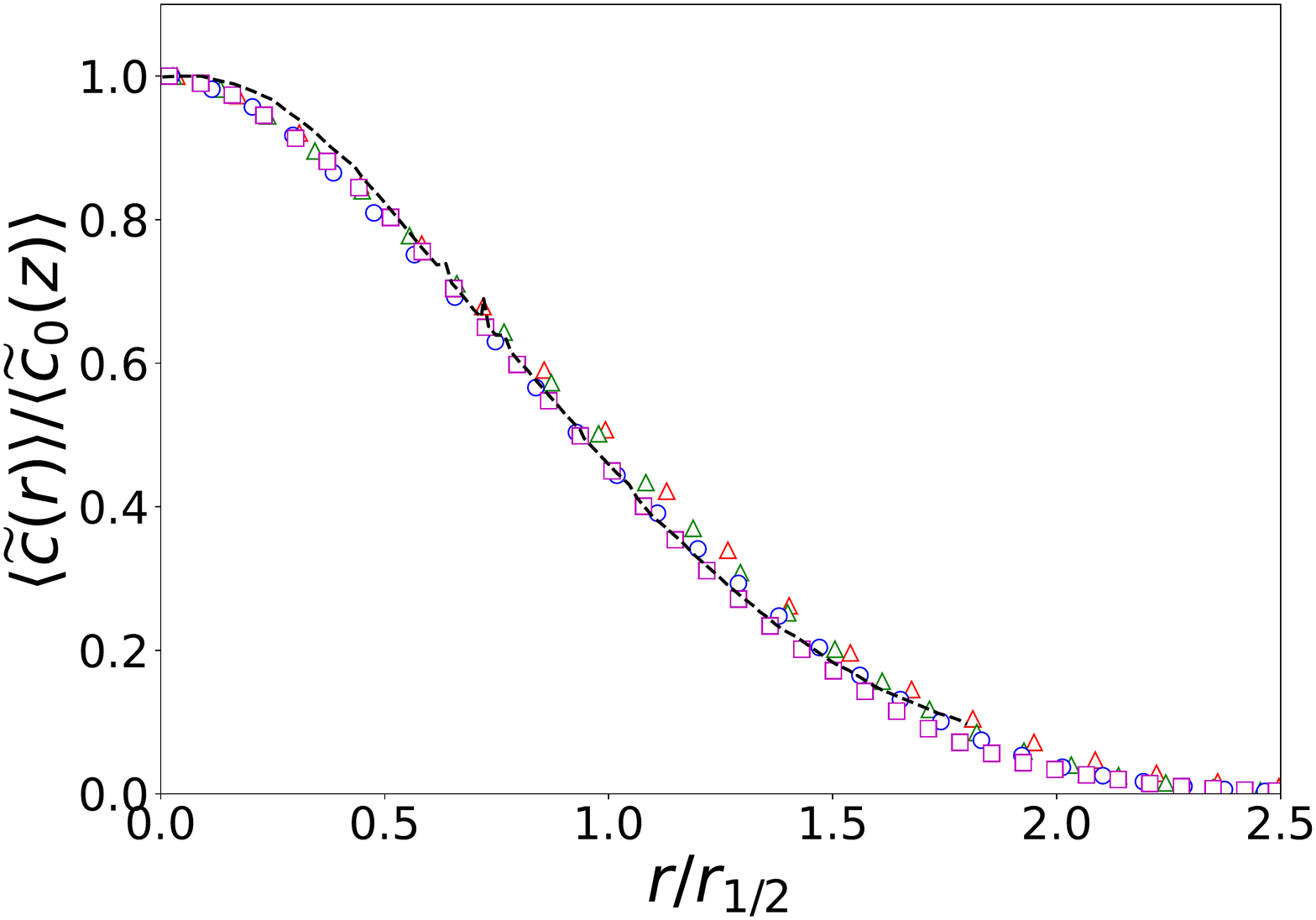}}
\caption{\protect\subref{fig:ss_vel} Axial mean velocity profiles as function of normalized radial distance, \protect\subref{fig:ss_conc} Mean {total} concentration profiles at $z/D_{J} = 161$ (\protect\rtri), $z/D_{J} = 213$ (\protect\gcircle), $z/D_{J} = 252$ (\protect\bsq) and $z/D_{J} = 342$ ($\mtri$) as a function of self similarity variable $r/r_{1/2}$. {The total concentration refers to 
$\tilde{c} = \sum{\tilde{c}_i}$, where $\tilde{c}_i = \tilde{n}_i \frac{\pi}{6}d_i^3$.}
The dashed line (\protect \blackldline) denotes the DNS data \citep{Lubbers2001}.}
\label{fig:sim_prof}

\end{figure*}

\begin{figure}
    \centering
    \includegraphics[width = 0.8\columnwidth]{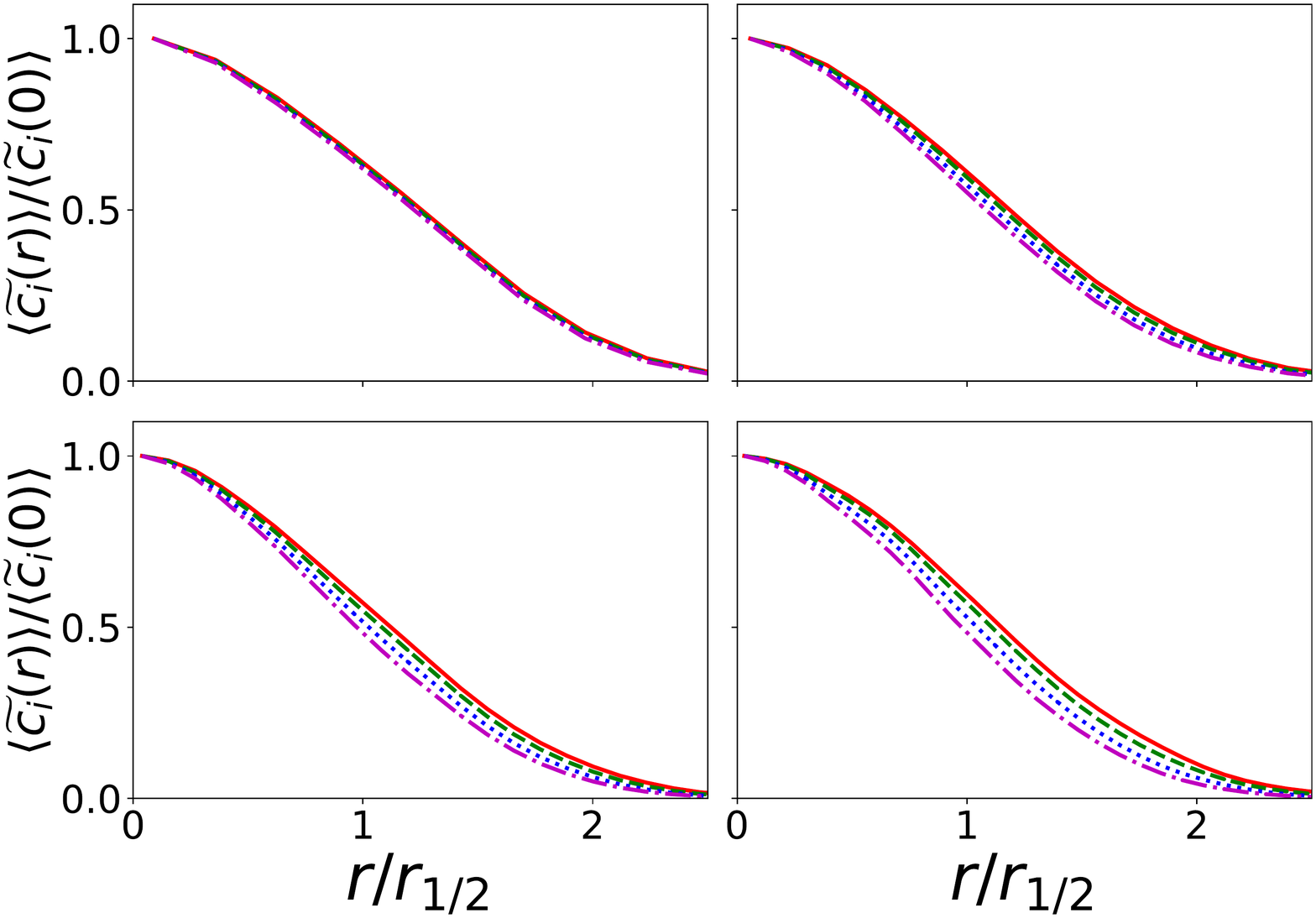}
    \caption{Normalized radial concentration profiles for different droplet sizes at (a) $z/D_J =108$, (b) $z/D_J = 160$, (c) $z/D_J = 212$, (d) $z/D_J = 264$ . The lines correspond to the individual droplets of diameter  $d = 50\;\mu m$ (\protect\redline), $d = 366\;\mu m$ (\protect\greendline), $d = 683\;\mu m$ (\protect\blueddline) and $d = 1\;mm$ ($\protect\magdline$). }
    \label{fig:conc_diff_plumes}
\end{figure}

\begin{figure}
    \centering
    \includegraphics[width = 0.8\columnwidth]{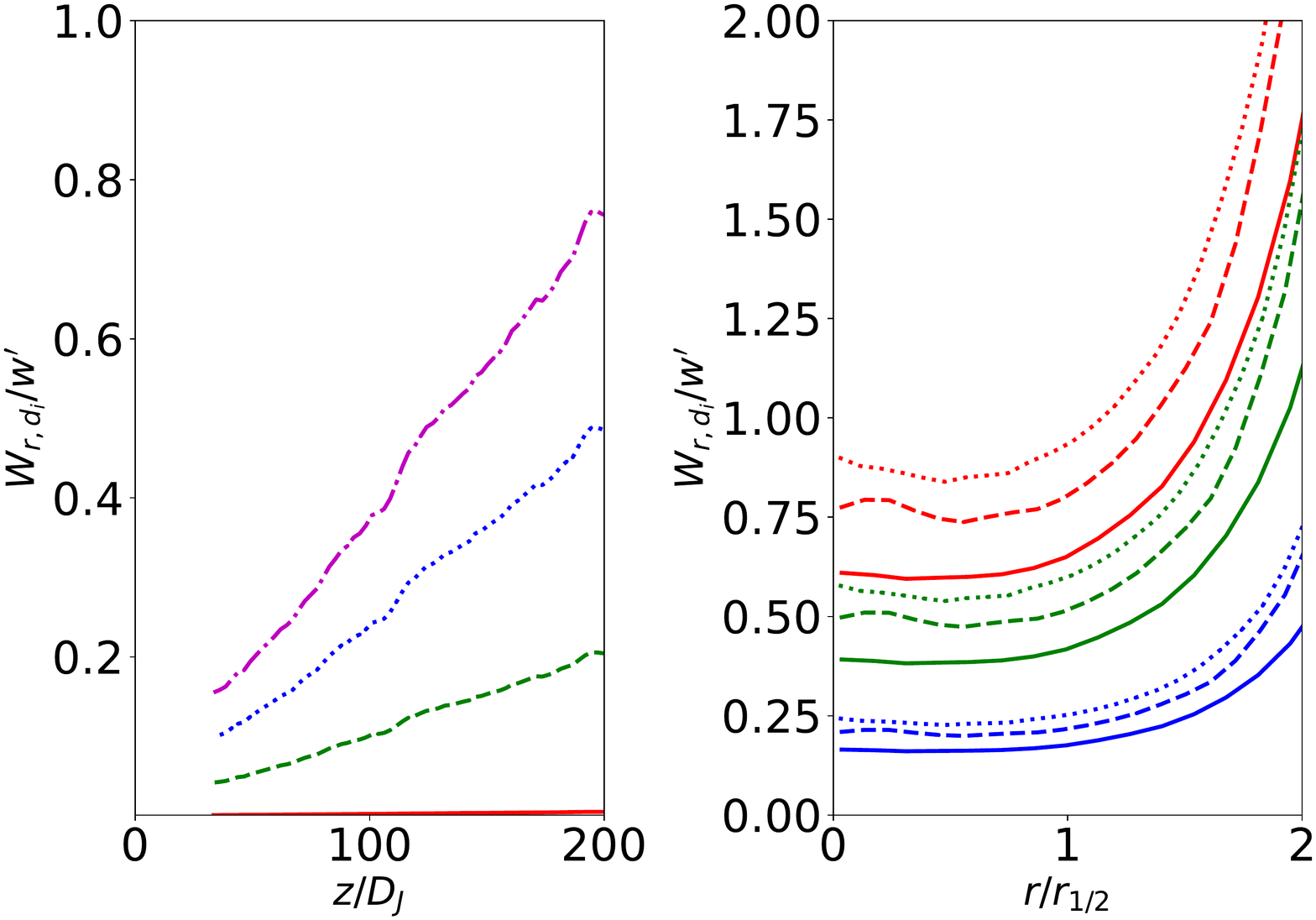}
    \caption{{ (a) Axial and (b) radial evolution of the settling parameter $Sv = W_{r,d_i}/w'$  for $d = 50\;\mu m$ (\protect\redline), $d = 366\;\mu m$ (\protect\greendline), $d = 683\;\mu m$ (\protect\blueddline) and $d = 1\;mm$ ($\protect\magdline$). In the right panel the radial profiles for each droplet size are shown at three downstream locatoins, $z/D_J = $, $z/D_J = $, and $z/D_J = $.} }
    \label{fig:slip_vel}
\end{figure}

Results for the   radial profiles of the mean velocity and total concentration at various downstream locations are shown in Fig. \ref{fig:sim_prof}. We see that the profiles show good collapse when plotted as a function of the {scaled} coordinate $r/r_{1/2}$. The LES also shows relatively good agreement when compared to DNS data for passive scalars from \citet{Lubbers2001}. {Profiles for the half-width and the vertical velocity for the jet for a similar configuration have been previously  discussed in \citet{Aiyer2020CoupledJet} and have been omitted here for the sake of brevity.}
The profiles of the individual concentration fields are shown in Figure \ref{fig:conc_diff_plumes} for four representative sizes and at different downstream locations. At the location nearest to the nozzle, {at} $z/D_J = 108$, we see that there is little difference between the radial profiles for the different droplet sizes. This is in agreement with the qualitative trends seen in the contour plot shown in Figure \ref{fig:inst_profiles}. Further downstream, we can see an increasing size dependence in the radial distribution of the concentration field where the smaller droplet sizes have a wider width as compared to the larger droplets. This observation is consistent with \citet{Gopalan2008a} and can be attributed to the increasing ratio of rise velocity to the turbulent axial fluctuating velocity component, {$S_v = W_{r,d_i}/w^{\prime}$, where $S_v$ is the settling parameter \citep{Chamecki2019}. The axial and radial profiles of the settling parameter for different droplet sizes are shown in Figure \ref{fig:slip_vel}.}   When the ratio increases, droplets  move from one eddy to another faster than the eddy decay rate.  For instance, nearer to the nozzle at $z/D_J = 108$, $W_{r,d_{10}}^2/w^{\prime 2} = 0.16$, and the ratio of the transverse particle to fluid diffusion coefficient can be estimated using Eq. (\ref{eqn:dispersion_coeff}) to be $D_{p,T}/D_{f,T} = 0.8$. Further downstream at $z/D_J = 200$, $W_{r,d_{10}}^2/w^{\prime 2} = 0.64$ and $D_{p,T}/D_{f,T} = 0.56$. {We can see from Figure \ref{fig:slip_vel}b that the ratio also increases as a function of radial distance as the turbulence gets weaker towards the edge of the jet} This results in decreased dispersion for the larger droplets as a function of axial and radial distance.

\begin{figure}
    \centering
       \subfloat[{Concentration}\label{fig:inv_conc}] {\includegraphics[width = 0.48\columnwidth]{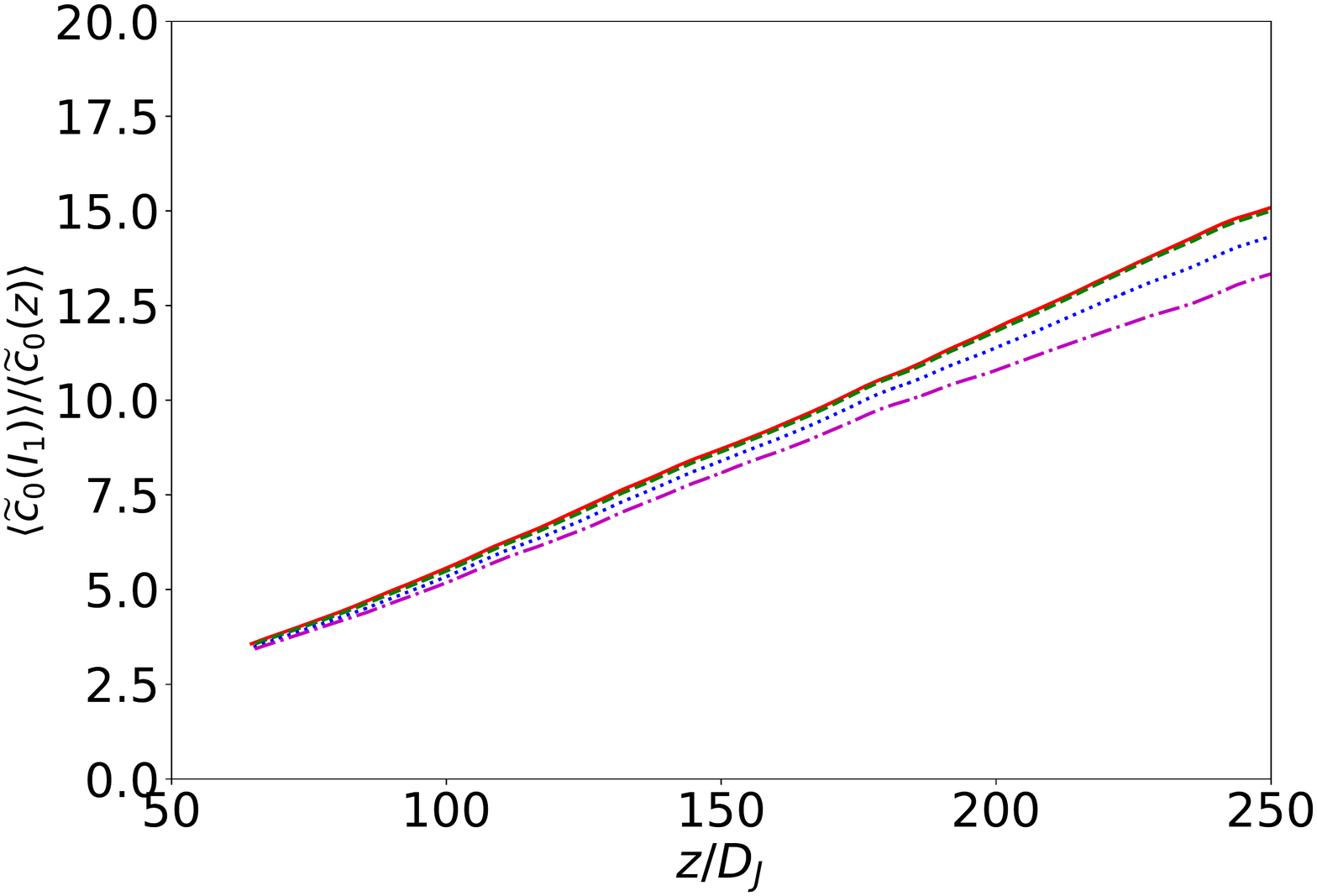}}
       \hfill
   \subfloat[Half-Width\label{fig:r_half_conc}] {\includegraphics[width = 0.48\columnwidth]{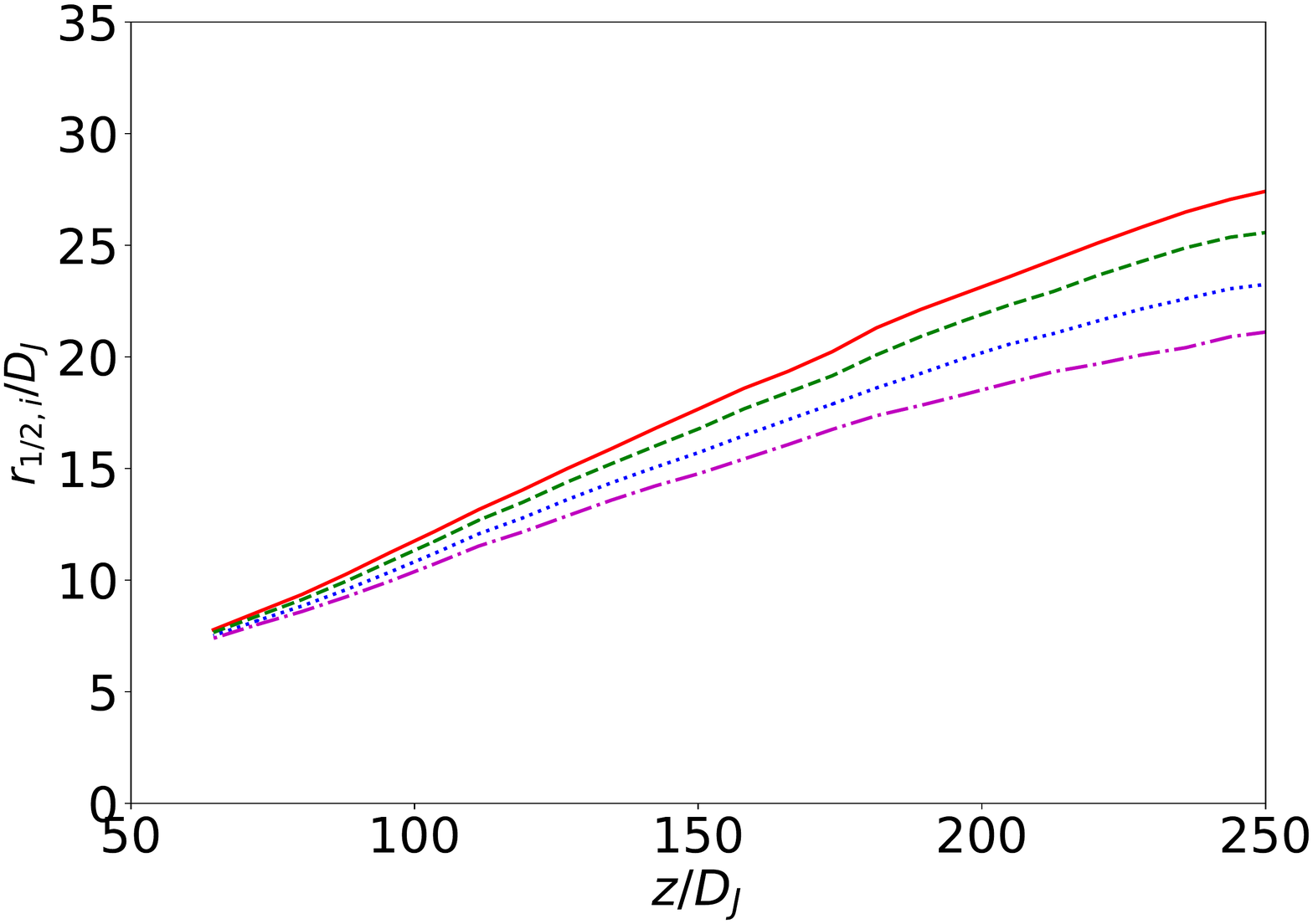}}
  
    \caption{{ Evolution of (a) inverse centerline concentration and }(b) concentration half-width as a function of downstream distance.  The lines correspond to the individual droplets of diameter  $d = 50\;\mu m$ (\protect\redline), $d = 366\;\mu m$ (\protect\greendline), $d = 683\;\mu m$ (\protect\blueddline) and $d = 1\;mm$ ($\protect\magdline$). }
    \label{fig:inv_conc_r_half}
\end{figure}

In Figure \ref{fig:inv_conc_r_half} we show the evolution of the {inverse centreline concentration and the} half-width as a function of axial distance. The  half-width is defined as usual as the radial location at which the concentration decays to half its centreline value:
    \begin{equation}\label{eqn:half_width_conc}
    \langle \tilde{c}_i(r=r_{1/2,i},z) \rangle = \frac{1}{2}\langle\tilde{c}_{0,i}(z)\rangle,
\end{equation}
where $c_i$ is the droplet concentration  in the ith bin and $r_{1/2,i}$ is the half width of the concentration in that bin. 
Moving downstream from the nozzle, the half-width profiles diverge for different droplet sizes, with the largest droplet having the smallest slope. The growth is initially linear for all the sizes and then curves, appearing to saturate for the larger droplets as a function of downstream distance. This suggests that a simple linear growth of the concentration half-width, i.e 
$d r_{1/2}/d z = S_d$ is only accurate for droplets with a small rise velocity {compared to the  root-mean-square of turbulence velocity fluctuations.} {The inverse concentration in Figure \ref{fig:inv_conc} shows similar size-dependent behaviour where the growth is smallest for the $d=1$mm droplet.}

\begin{figure}
    \centering
    \includegraphics[width = 0.8\columnwidth]{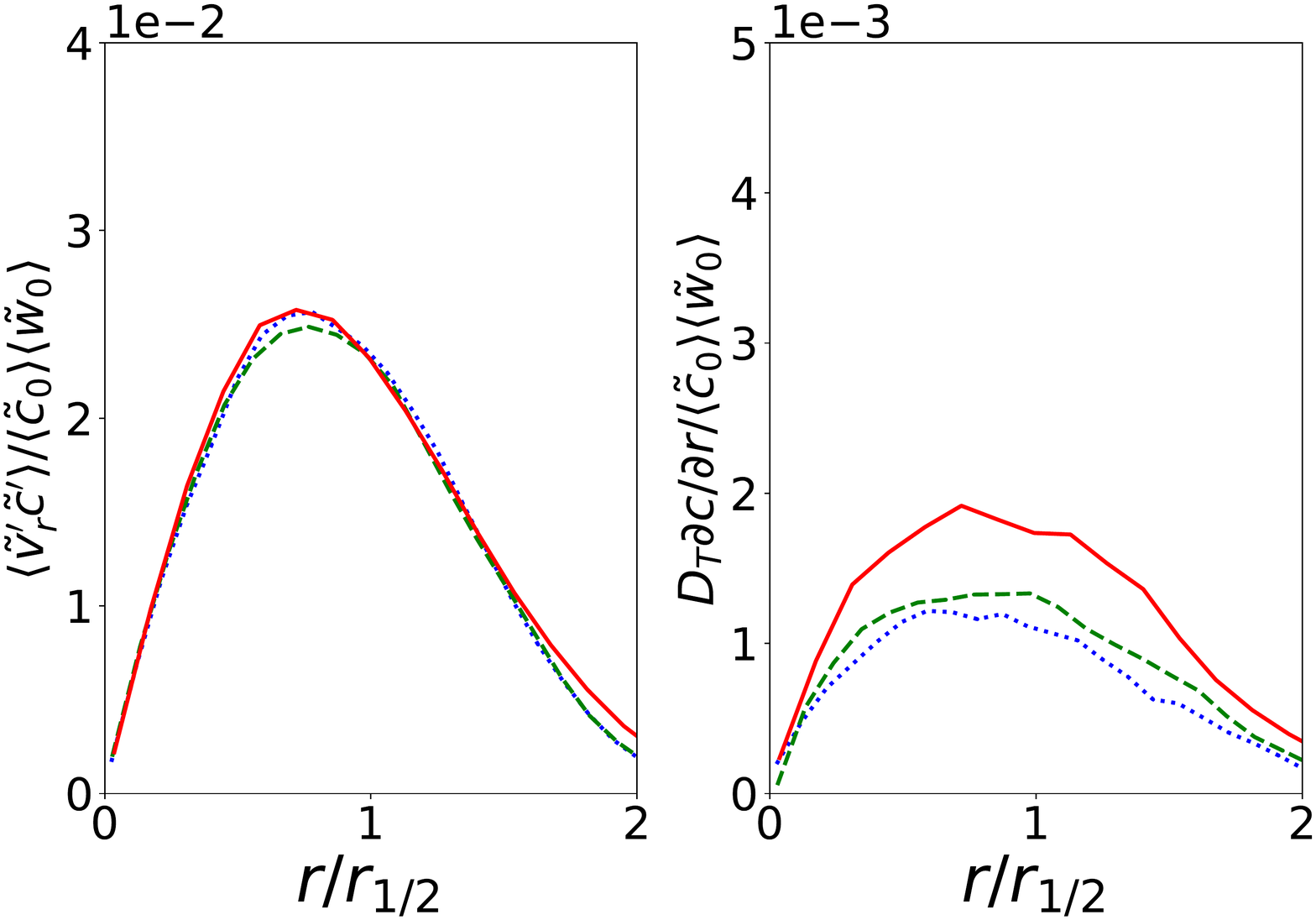}
    \caption{a) Total normalized radial turbulent concentration flux and b) subgrid flux contribution at $z/D_J = 160$ (\protect\redline), $z/D_J = 212$ (\protect\greendline) and $z/D_J = 238$ (\protect\blueddline) as a function of self similarity variable $r/r_{1/2}$ . }
    \label{fig:conc_flux_tot}
\end{figure}

The concentration flux $\langle u_i^{\prime}{c}^{\prime}\rangle$ quantifies the efficiency of turbulent transport affecting the concentration field. We plot the radial resolved {total (i.e. including all sizes) concentration} flux $\langle\tilde{v}_r^{\prime}\tilde{c}^{\prime}\rangle$  at different downstream locations in Figure \ref{fig:conc_flux_tot}. The flux is normalized with the centreline vertical velocity and concentration. We can see that the resolved total concentration flux exhibits a self-similar behaviour as the profiles at various downstream distances collapse when plotted as a function of the {scaled} coordinate. The subgrid concentration flux is shown in Figure \ref{fig:conc_flux_tot}b.  The subgrid flux is defined as  $-\langle D_{sgs} \partial \tilde{c}/\partial r\rangle $, where $D_{sgs} = \nu_{sgs}/Sc_{sgs}$ is the turbulent SGS eddy diffusivity for the concentration field. The maximum contribution of the subgrid flux ($\approx 8 \%$) is near the source of the jet where the length scales of the flow are comparable to the grid resolution. Further downstream the subgrid flux reduces to less than $5 \%$ of the resolved concentration flux. The lower contribution from the subgrid model is expected as the the length scales of the jet grow as a function of downstream distance to the nozzle, and the grid resolution becomes sufficient to resolve the majority of the flux. Coarser grid resolutions are expected to lead to a higher subgrid contribution to the total flux. We remark that in cases of LES using very coarse grids, the constant Schmidt number SGS parameterization may be insufficient to capture the size based dispersion characteristics.

\begin{figure}
    \centering
    \subfloat[$z/D_J = 150$]{
    \includegraphics[width = 0.8\columnwidth]{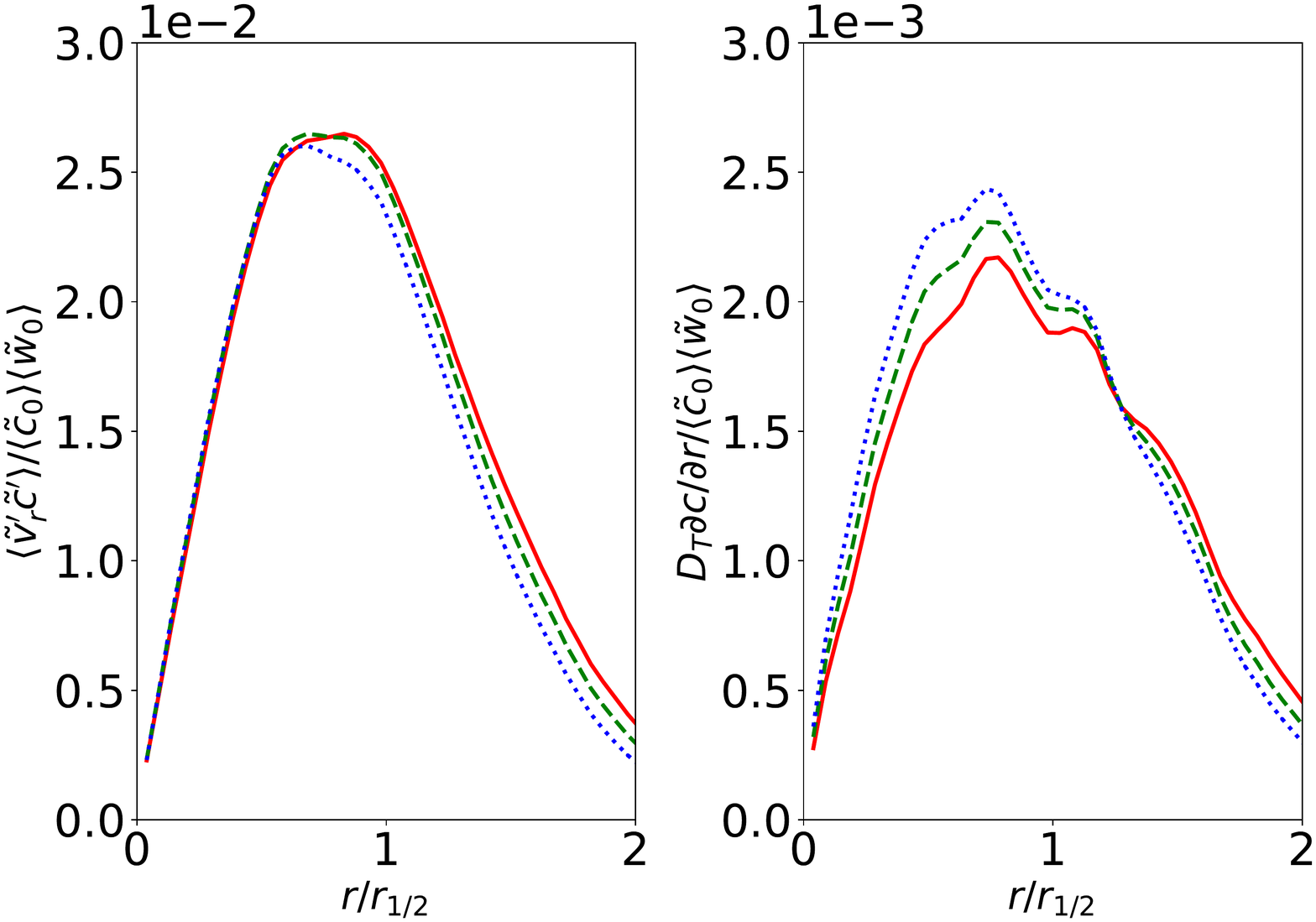}}
    \hfill
    \subfloat[$z/D_J = 200$]{
    \includegraphics[width = 0.8\columnwidth]{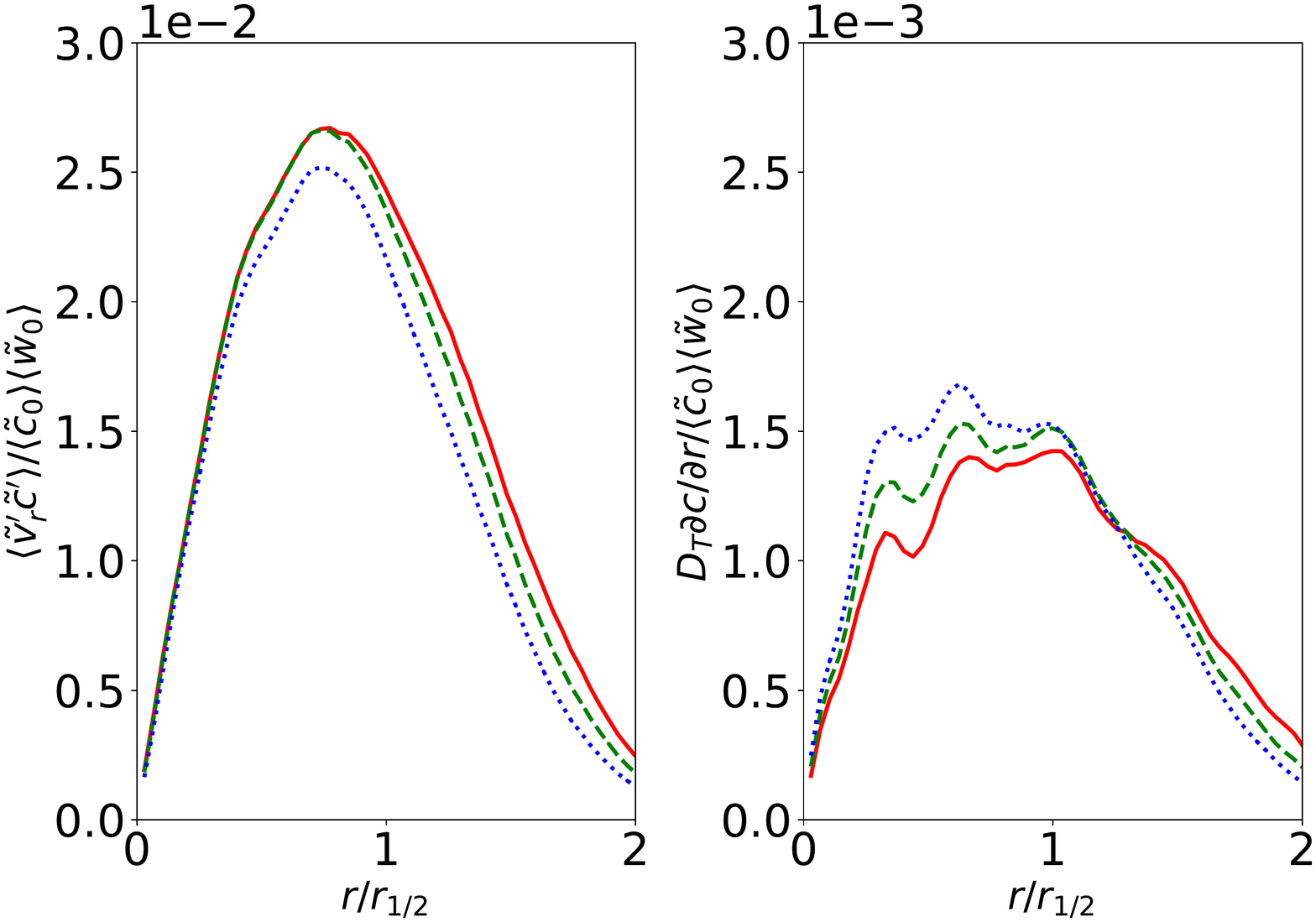}}
    \caption{Normalized radial turbulent concentration flux and subgrid flux contribution at a) $z/D_J = 150$ and b) $z/D_J = 200$ as a function of self similarity variable $r/r_{1/2}$. The symbols are  $d = 366\ \mu m$ (\protect\redline), $d = 683\ \mu m$ (\protect\greendline) and $d = 1000\ \mu m$ (\protect\blueddline) . }
    \label{fig:conc_flux_in}
\end{figure}

The transport of individual droplet sizes $\langle\tilde{v_r}^{\prime}\tilde{c}_i^{\prime}\rangle$ is examined in Figure \ref{fig:conc_flux_in} at $z/D_J =150$ and $z/D_J = 200$. Nearer to the nozzle the difference in the resolved fluxes is small, with the smaller droplet size displaying a larger flux. The difference in the flux increases further from the nozzle. We see that the concentration field of droplets  with diameter $d = 366\ \mu m$ has a maximum normalized flux that is higher than that for the largest droplets with $d = 1\ mm$ at $z/D_J = 200$. At this location, the differential dispersion has become more apparent, affecting the transport of the different droplet sizes. The smaller droplets are transported more efficiently by the flow than the larger ones resulting in larger transverse dispersion. Note that for the high resolution LES considered, we use a constant Schmidt number for the SGS concentration flux and the resolved scales are primarily responsible for the differential dispersion as the subgrid contribution to the flux is always $< 10\%$. We can conclude that an Eulerian-Eulerian LES model with the droplet velocity modelled as a function of the droplet time-scale accurately captures crossing trajectory  effects when the grid resolution is sufficiently high.

\subsection{Modified Schmidt number}
\begin{figure}
    \centering
     \subfloat[Droplet half-width \label{fig:rhalf_diff_plumes}]
    {\includegraphics[width = 0.48\columnwidth]{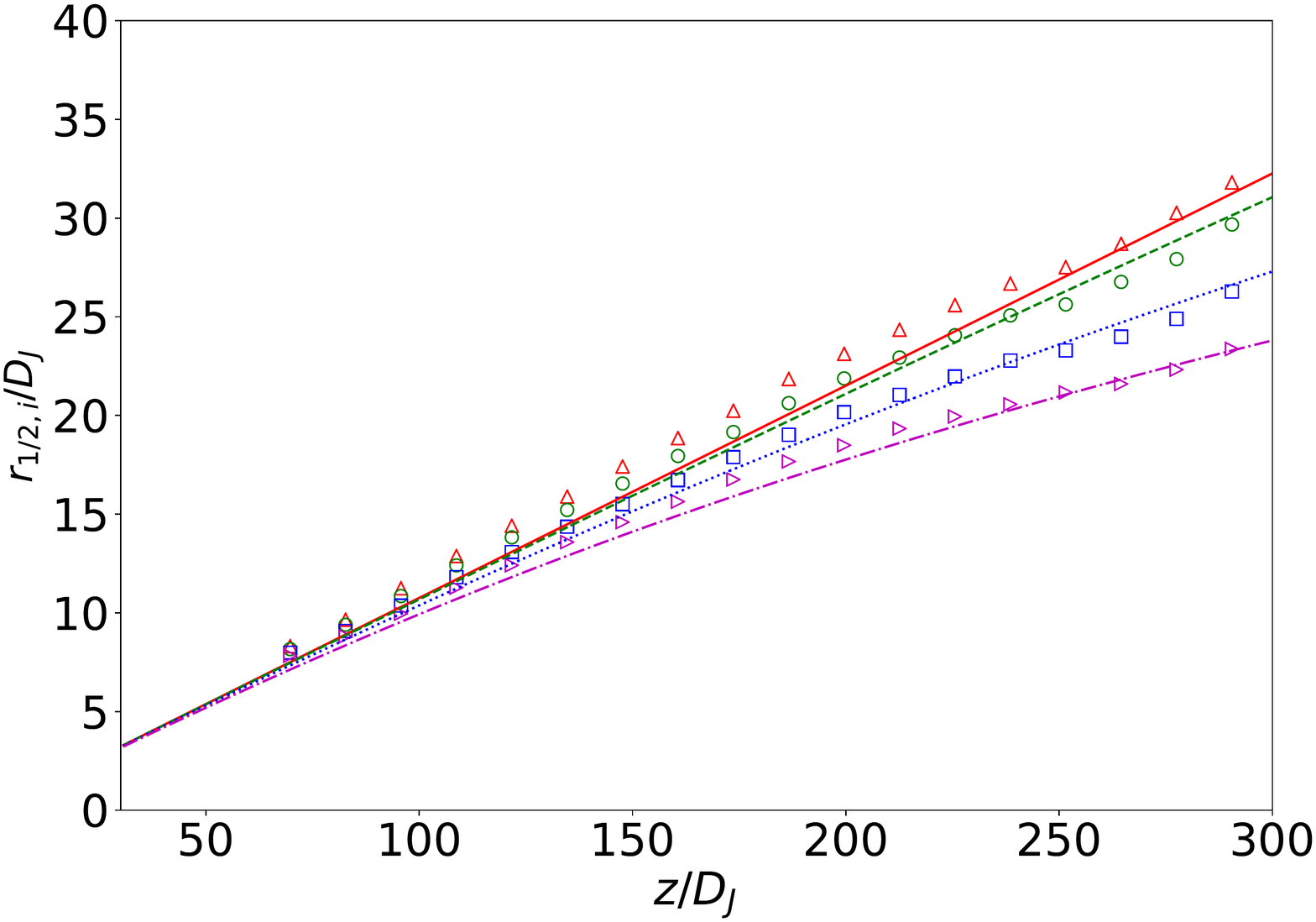}}
    \hfill
    \subfloat[Transverse diffusion coefficient \label{fig:transvers_coeff}]
    {\includegraphics[width = 0.48\columnwidth]{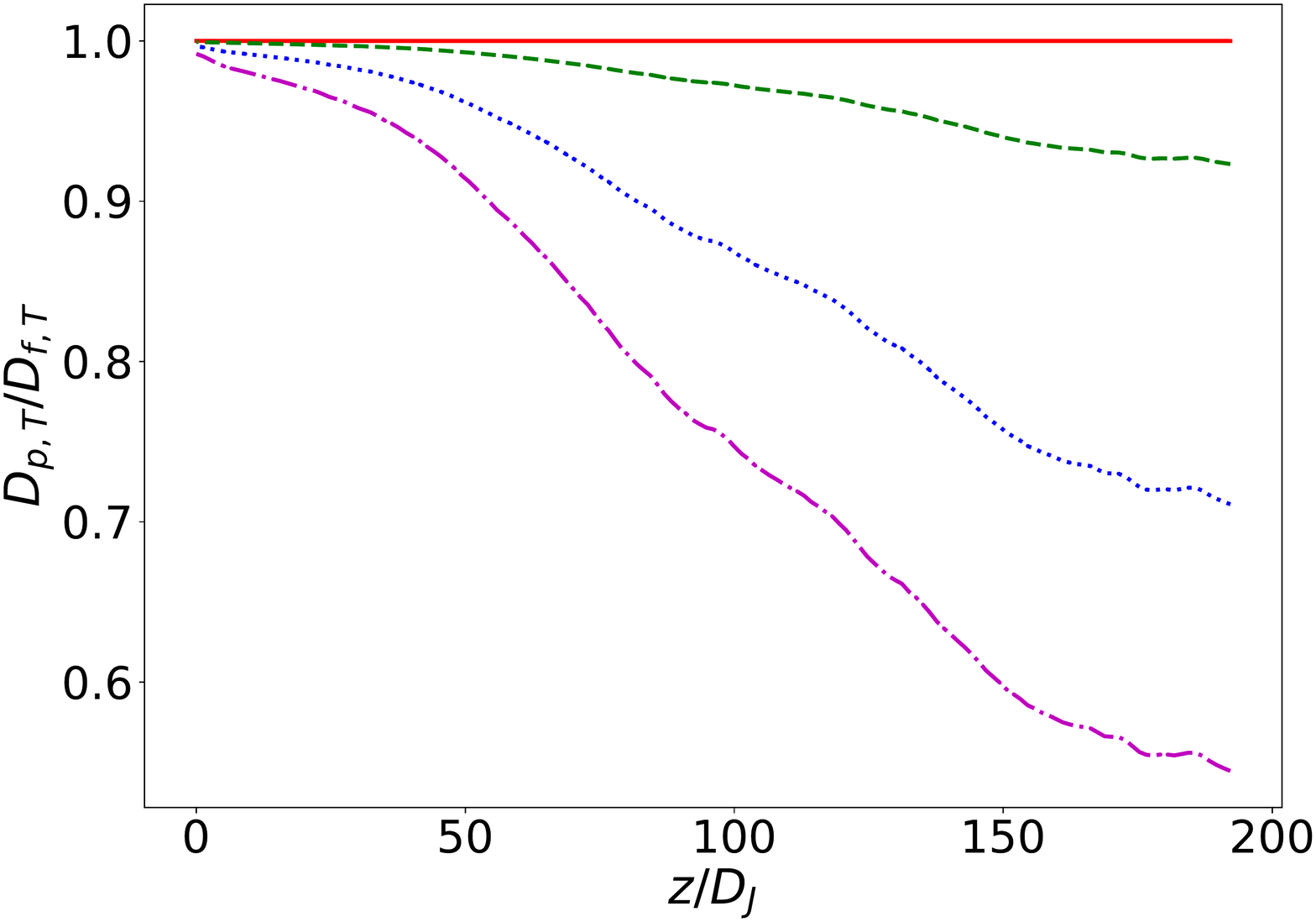}}
    
    \caption{\subref{fig:rhalf_diff_plumes} Comparison of concentration half width of different droplet sizes from LES (symbols) and the model (Eq. (\ref{eqn:rhalf_model})) \subref{fig:transvers_coeff}) Transverse diffusion coefficient as defined by Eq. (\ref{eqn:dispersion_coeff}) as a function of axial distance.  The 
    lines and the corresponding color coded symbols correspond to the individual droplets of diameter  $d = 50\;\mu m$ 
    (\protect\redline, \protect\rtri), $d = 366\;\mu m$ 
    (\protect\greendline, \protect\gcircle), $d = 683\;\mu m$ (\protect\blueddline, 
    \protect\bsq) and $d = 1\;mm$ (\protect\magdline, 
    $\protect\mtri$). }
    \label{fig:rh_trans}
\end{figure}

 We now examine further Eq. (\ref{eqn:dispersion_coeff}) proposed by \citet{Csanady1963} in order to construct a model to capture the differential size-based spread of the droplet plumes. We focus on a turbulent jet where the jet characteristics, such as downstream centreline velocity, half-width and centreline dissipation can be described by well known parameterizations. In Eq. (\ref{eqn:dispersion_coeff}), we use the droplet rise velocity  defined in Eq. (\ref{eqn:rise_vel}). The mean turbulent intensity in the direction of the rise velocity can be estimated based on the jet centreline velocity as $w' \approx 0.35 w_0(z)$ \citep{Hussein1994} and $w_0(z) = C_u D_J w_0/z$ with $C_u \approx 6$. We plot the ratio of the transverse dispersion coefficient to the fluid's turbulent dispersion for four different droplet sizes in Figure \ref{fig:transvers_coeff}. We can see that the ratio is near unity for the smaller droplets and reduces for larger droplet sizes as a function of downstream distance. Further from the jet source, the mean velocity and turbulence intensity decreases, resulting in an increase in $W_r/w'$ thereby reducing the dispersion coefficient. We observe the same behaviour qualitatively in the LES, where the dispersion of larger droplets that have a higher rise velocity is suppressed.

The similarity solution for the concentration profile based on a constant eddy-diffusivity hypothesis is given by \citep{Law2006, Aiyer2020CoupledJet}:
 \begin{equation}\label{eqn:sim_conc_ch_6}
     \overbar{c}(r,z) = \frac{\overbar{c}_0(z)}{\left(1+\alpha^2\eta^2\right)^{2 Sc_T}},
 \end{equation}
where $\alpha^2 = (\sqrt{2}-1)/S^2$ with $S$ being the spread rate for the mean velocity field, $Sc_T$ is the turbulent Schmidt number and $\eta = r/z$ is the similarity variable. The derived solution does not incorporate size based effects in the mean profile. One can account for the effect of finite particle rise velocity in Eq. (\ref{eqn:sim_conc_ch_6}) by allowing the turbulent Schmidt number $Sc_T$ to be size dependent. This size dependence 
can be incorporated through an expression based on the transverse diffusion coefficient as expressed in Eq. (\ref{eqn:dispersion_coeff}).  The corrected Schmidt number is then given by:
\begin{equation}\label{eqn:Sc_correction}
    Sc_T(z,d_i) = Sc(0) \, \sqrt{1 + 4\,C\, \frac{W_{r,d_i}^2}{w^{\prime 2}(z)} } \, ,
\end{equation}
where $Sc(0)$ is the size independent baseline coefficient and  $w'$ is the root-mean-square (r.m.s.)  of axial turbulent velocity fluctuations.  $w'$ is estimated from the centreline axial velocity using   $w^{\prime} = 0.35 \, \overbar{w}_0(z)$ \citep{Hussein1994}. {The constant $C$ is the ratio of the Lagrangian to Eulerian timescale. \citet{Yamamoto1987} measured the timescale ratio experimentally to be between $0.3-0.6$. \citet{Corrsin1963} developed an approximate  relation between the Lagrangian and Eulerian timescales of fluid points which was a function of the fluid turbulence, $C= T_L/T_E =  A \langle w \rangle /w'$ where the constant $A$ is in a range $0.3 - 0.8$. Estimating $w' = 0.35 \langle w \rangle$ at the centerline of the jet, we obtain that $C$ is in a range 0.85-2. In this work we select $C=1$ for simplicity.} The half-width of the concentration profile can be calculated from Eq.  (\ref{eqn:sim_conc_ch_6}) and (\ref{eqn:Sc_correction}). At $r=r_{1/2}$,  $c(r,z)/c_0(z) = 1/2$,  and we obtain
\begin{equation}\label{eqn:int_rhalf}
    \frac{1}{2} = \frac{1}{\left(1 + \alpha^2 \, r_{1/2}^2/z^2\right)^{2 \,Sc_T}}.
\end{equation}
Rearranging Eq. (\ref{eqn:int_rhalf}) we can derive the evolution of the half-width as a function of downstream distance as :
\begin{equation}\label{eqn:rhalf_model}
    r_{1/2} (z,d_i) = 
    \left[\frac{1}{\alpha}\left({2}^{1/({2Sc_T})} - 1\right)^{1/2} \right]\,\,z,
\end{equation}
For a constant $Sc_T$, Eq. (\ref{eqn:rhalf_model}) recovers the usual linear dependence of the half-width on axial distance with a slope equal to $1.14 \, S$ (for $Sc_T=0.8$, which is typical for turbulent round jets \citep{Chua1990}), where $S$ is the spread rate for the velocity field. Eq. (\ref{eqn:rhalf_model}) suggests that for droplets with finite rise velocity, the slope is no longer a constant but  depends on $z$ via the Schmidt number (that appears in the exponent in Eq. \ref{eqn:rhalf_model}) dependence on $z$ (Eq. \ref{eqn:Sc_correction}). 

The half-width calculated using the modified Schmidt number and the derived similarity profile can be validated with that calculated directly from the concentration profiles of the various droplet plumes from the LES. The Schmidt number is calculated using Eq. (\ref{eqn:Sc_correction}) with the typical value of $Sc(0) = 0.8$ \citep{Chua1990}.  We can see from Figure \ref{fig:rhalf_diff_plumes} that the modified Schmidt number approach predicts the spread of the droplet plumes for different sizes with good accuracy, even though the analytical solution on which the expression is based (Eq. \ref{eqn:sim_conc_ch_6}) was derived for a constant $Sc_T$. Using the local $Sc_T(z,d_i)$ in the constant $Sc_T$ solution must be considered an approximation, but results show that it captures rather well the size dependence associated with the profiles of the concentration with a larger rise velocity. In the next section, we describe how this approach can be adapted as a subgrid model for the concentration flux in coarse large eddy simulations.

\section{A modified Schmidt number SGS model}\label{sec:subgrid}
 In the previous section we demonstrated that with sufficient grid resolution, an SGS model where the subgrid concentration flux was parameterized using a constant turbulent Schmidt number accurately characterized the size based differential dispersion of the droplets. For coarser simulations, the contribution of the subgrid flux would be more important, and a parameterization that incorporates dependence on the droplet size (or rise velocity) is needed.
  \begin{figure}
    \centering
    \includegraphics[width = 0.8\columnwidth]{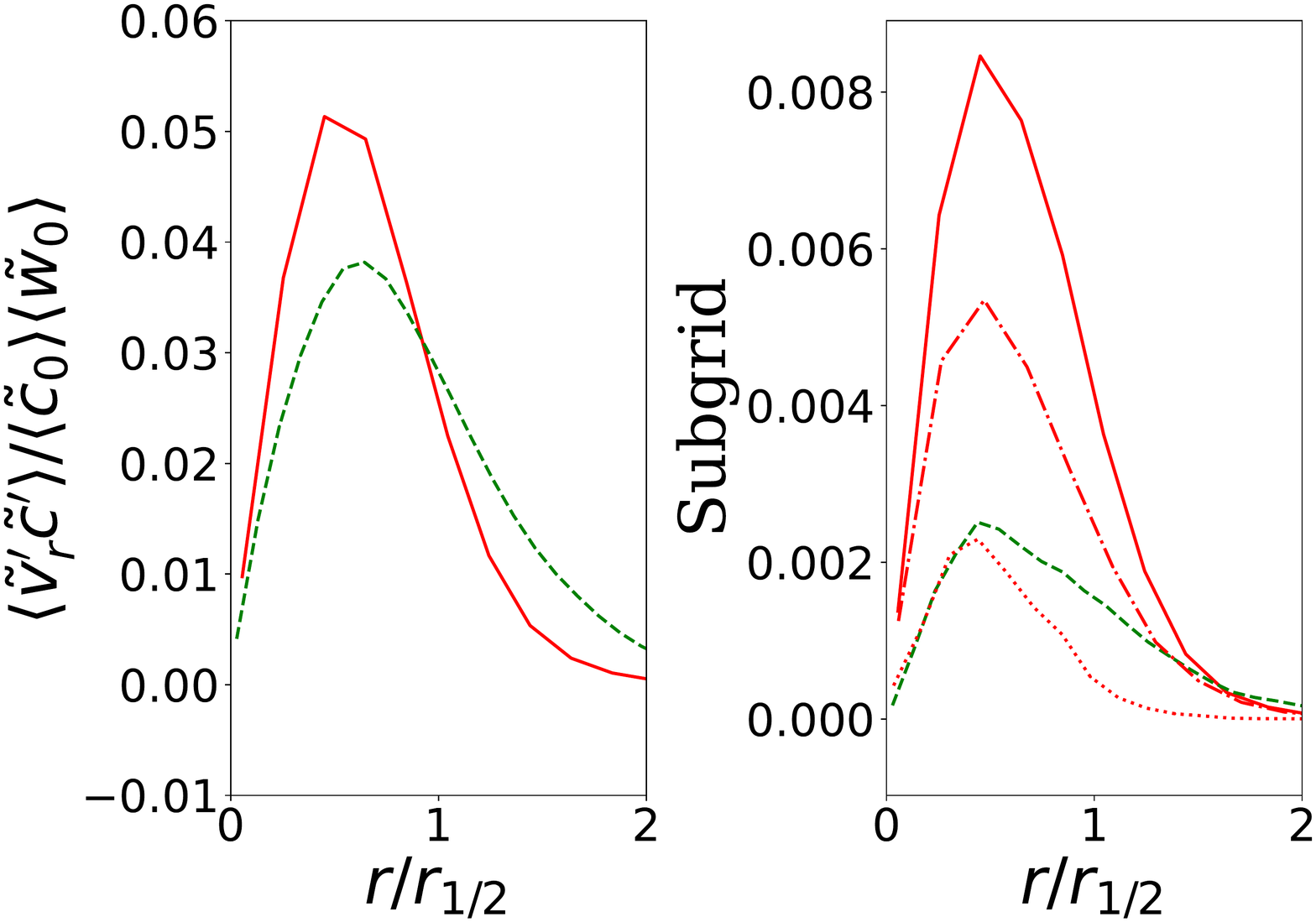}
    \caption{a) Normalized radial turbulent concentration flux and b) subgrid flux contribution at $z_s/D_J = 80$ (\protect\redline), $z_s/D_J = 115$ (\protect\greendline)  as a function of self similarity variable $r/r_{1/2}$ for CS1. {Additionally shown are subgrid fluxes for FS1 (\protect\redddline)  and CS2 (\protect\reddashdline) at $z_s/D_J = 80$}}
    \label{fig:conc_flux_tot_coarse}
\end{figure}
The subgrid-scale concentration flux, $\ppi_{i} = (\widetilde{\vv_i n_i} - \tvv_i\tn_i)$ is parameterized as $\pi_{n,i} = -(\nu_{sgs}/Sc_{sgs})\nabla\tn_{i}$ where so far $Sc_{sgs}$ has been taken to be a constant. We now introduce a size dependent formulation for the Schmidt number by expressing $Sc_{sgs}$ in terms of the physics contained in Eq. (\ref{eqn:Sc_correction}). In   LES however, no information is available regarding the direction of unresolved subgrid scale eddies that interact with the rising particles. Hence,   no distinction can be made between the transverse and longitudinal directions in parameterizing the eddy diffusivity and some average behavior must be invoked for simplicity.  The corrected Schmidt number $Sc_{sgs}$ is thus calculated as the root mean square average value of the two expressions for longitudinal and transverse dispersion:
 \begin{equation}\label{eqn:Sc_sgs}
     Sc_{sgs}(\xx,t) = Sc_{sgs}(0) \, \sqrt{\frac{1}{2}\left(1 + \frac{C \, W_{r,d_i}^2}{w^{\prime 2}(\xx,t)} +1 + \frac{4 \, C\, W_{r,d_i}^2}{w^{\prime 2}(\xx,t)} \right)} =Sc_{sgs}(0) \, \sqrt{1 + \frac{2.5 \, W_{r,d_i}^2}{w^{\prime 2}(\xx,t)}} 
 \end{equation}
 where  $C = 1$ has been assumed and we use $Sc_{sgs}(0) = 0.4$ \citep{Yang2016}. {The ratio of the Eulerian to the Lagrangian timescale was set to unity since it is known to be of order 1 and no additional information is available to us in the context of LES.} {Moreover, note that in Equation \ref{eqn:Sc_sgs} the velocity $W_{r,d_i}$ stands for the magnitude of the relative velocity between phases. In the present application it is taken to be equal to the rise velocity but in more general settings could include other effects as well, such as the last term in Equation  \ref{eqn:rise_vel} \citep{Arcen2008}.}
 The local turbulence subgrid velocity fluctuation $w^{\prime}(\xx,t)$ is estimated in terms of the local  subgrid kinetic energy, the latter modeled in LES according to
\begin{equation}\label{eqn:wprime_mod}
    w^{\prime}(\xx,t) = \sqrt{\frac{2}{3}k_{sgs}(\xx,t) } , \,\,\,\,{\rm where}\,\,\,\, 
    k_{sgs}(\xx,t) = C_I \Delta^2 |\tilde{S}|^2(\xx,t)
\end{equation}
 using the SGS kinetic energy closure of \citet{Yoshizawa1991}.  The constant $C_I$ is approximated by $C_s^2$, the Smagorinsky coefficient determined from the LASD model (future model improvements to determine the combined $C$ and $C_I$ based on the Germano identity \citep{Germano91} could be envisioned), $\Delta$ is the filter scale and $|\tilde{S}|$ is the magnitude of the resolved strain-rate tensor. 

\begin{table}
\centering

\begin{tabular}{c c c  c  } 
 \centering No.
 & \multicolumn{1}{p{3cm}}{\centering 
 $\Delta x=\Delta y$} & $Sc_T $ &$\overbar{w}_{in}$ (m/s) \\ [5pt] \hline
FS1 & $3.4\times 10^{-3}$ & 0.4 & 2 \\ 
 CS1  & $1.7\times 10^{-3}$ & 0.4& 1 \\ 
 CS2  & $1.7\times 10^{-3}$& Eq. \ref{eqn:Sc_sgs} & 1 \\
\end{tabular}
\caption{Simulation parameters.}
\label{tab:Sim_param}
\end{table}

In order to test the modified Schmidt number SGS model, we perform two simulations with a resolution that is 4 times coarser than that described in section \S \ref{sec:Results}. The simulation details for the three cases are given in Table \ref{tab:Sim_param}. We use a constant SGS Schmidt number model ($Sc_{sgs}=0.4$) for simulation FS1 (fine-scale simulation) and CS1 (coarse simulation), while the modified Schmidt number model is used for CS2. The body force $\tilde{F}$ is applied 10 diameters ( $\Delta z = 10 D_J$) downstream of the fine simulation
(as described in section \S\ref{sec:Eul_sim}) and the corresponding injection velocity is set to match the velocity of simulation FS1 at $z = 20 D_J$ downstream of the true nozzle, as shown in Figure \ref{fig:sketch_sim}. In order to ensure the droplets in each simulation are injected into a LES velocity field that has the same mean velocity distribution, the droplet injection location is set further downstream, to $I_2 = 280 D_J$ as shown in Figure \ref{fig:sketch_sim}. Droplets of sizes $d = (50,\ 366,\ 683,\ 1000)\ \mu m $ are injected well within the self similar region of the jet, downstream of the momentum injection location.

\begin{figure}
        \centering
        \subfloat[$d = 50\ \mu m$\label{fig:d50}]{\includegraphics[width=0.48\columnwidth]{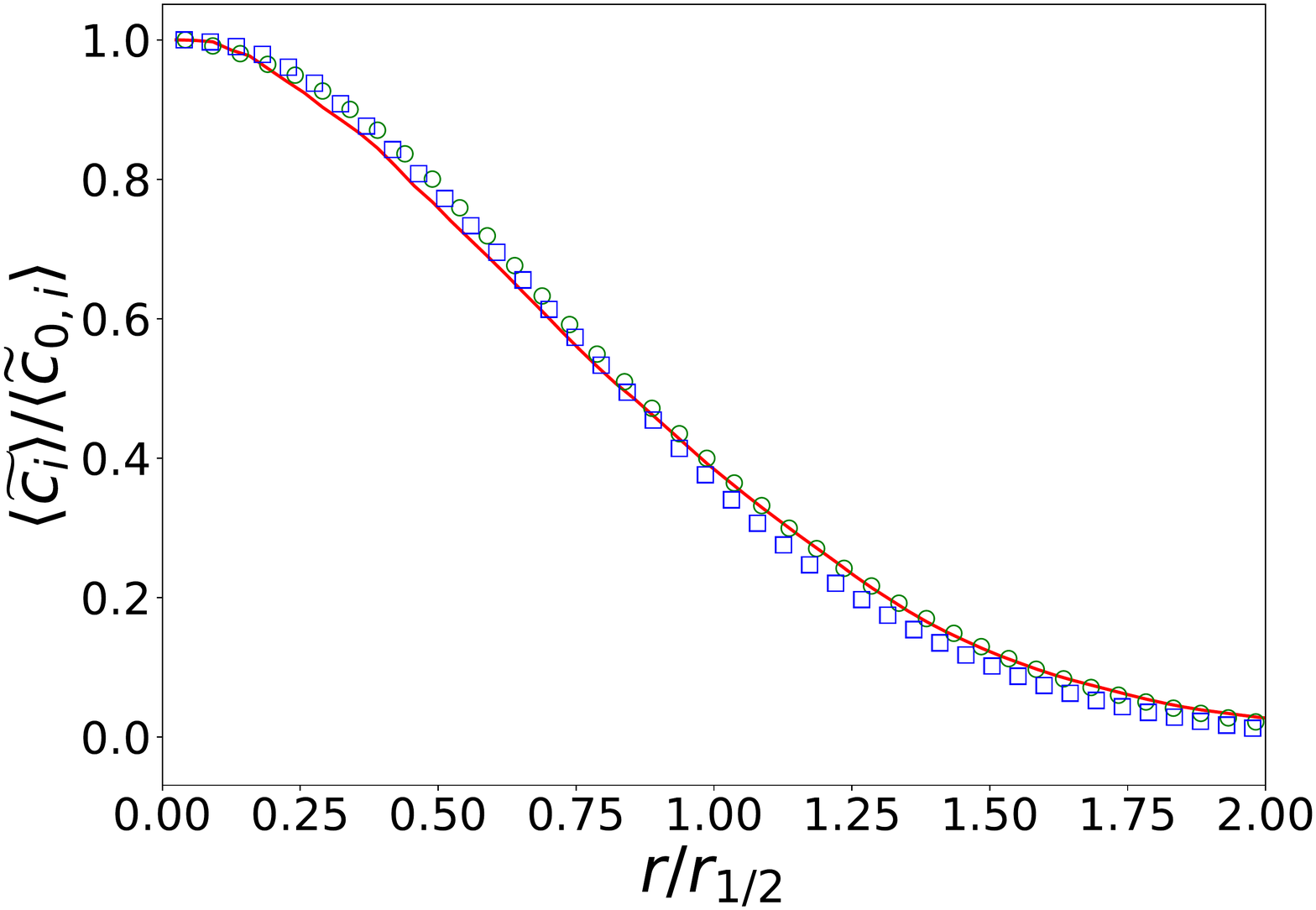}}
        \hfill
        \subfloat[$d = 360\ \mu m$ \label{fig:d360}]{\includegraphics[width=0.48\columnwidth]{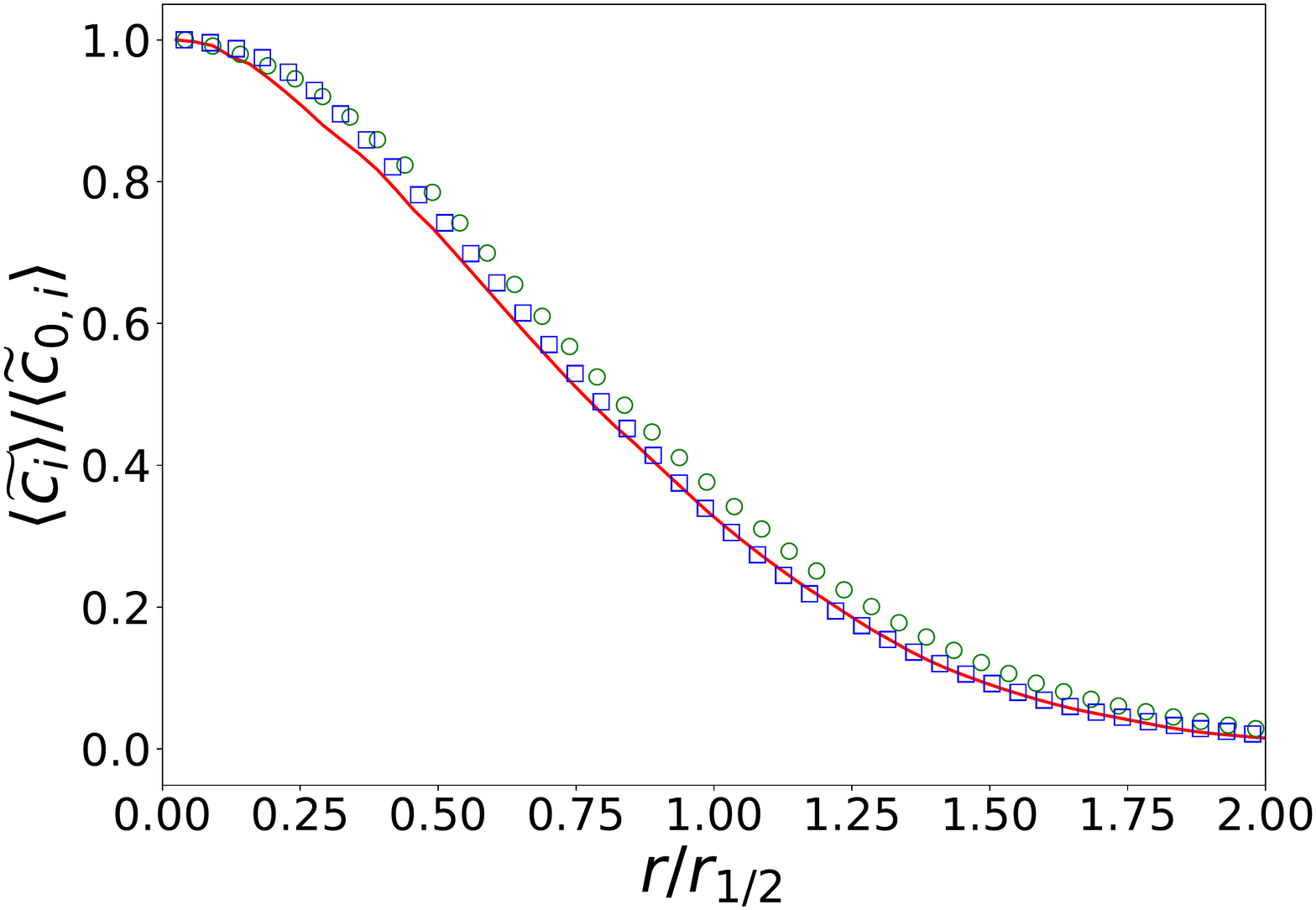}}\\
        \subfloat[$d = 683\ \mu m$ \label{fig:d683}]{\includegraphics[width=0.48\columnwidth]{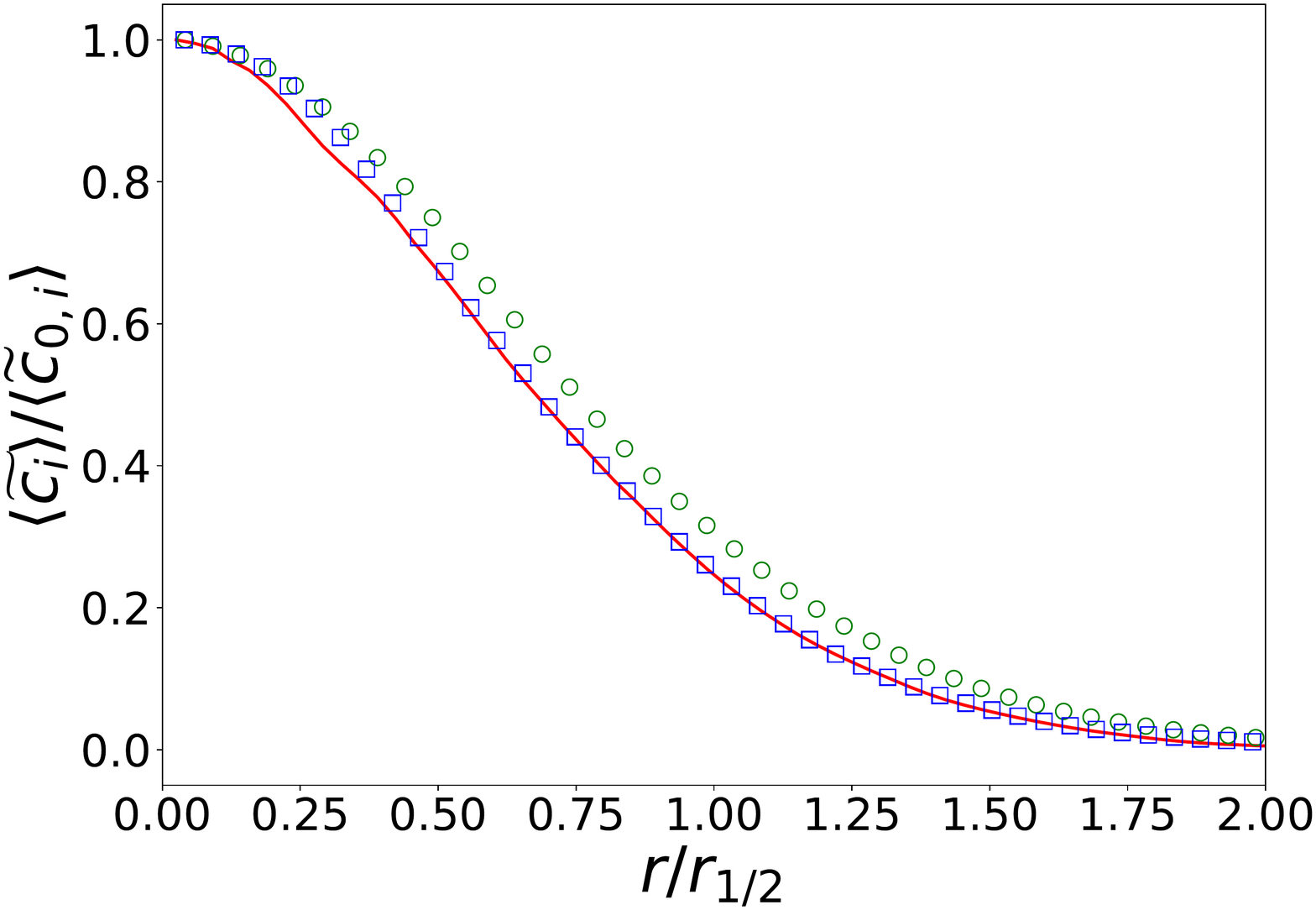}}
          \hfill
        \subfloat[$d = 1000\ \mu m$ \label{fig:d1000}]{\includegraphics[width=0.48\columnwidth]{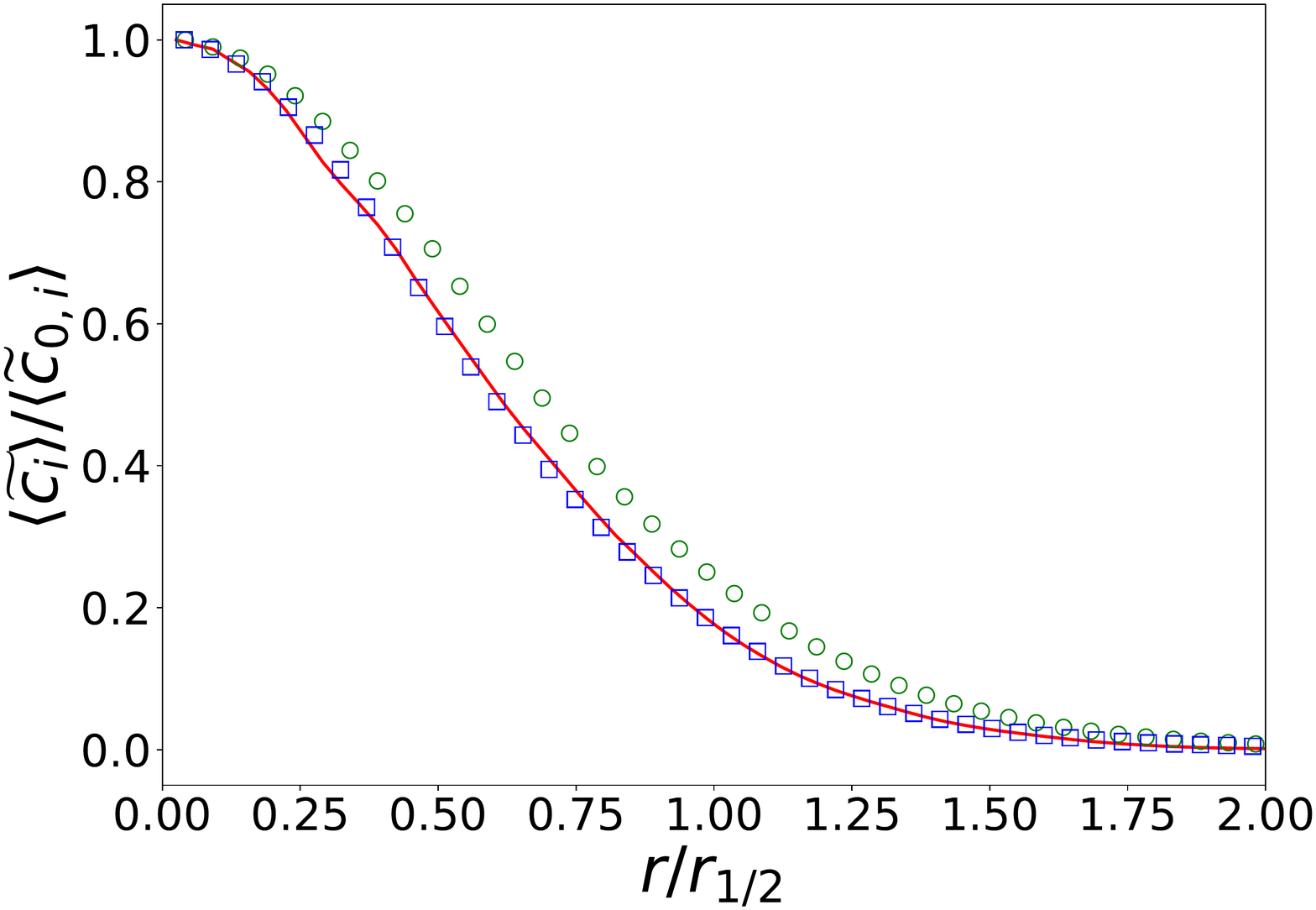}}
        
  \caption{Radial distribution of droplet concentration from the FS1, (\protect\redline) , CS1 (\protect\gcircle) and CS2 (\protect\bsq)  for \subref{fig:d50}) $d = 50 \mu m$, \subref{fig:d360}) $d = 360 \mu m$, \subref{fig:d683}) $d = 683 \mu m$, and \subref{fig:d1000}) $d = 1000 \mu m$ at $z_{s}/D_J = 169$.}
    \label{fig:conc_drop_sgs}
\end{figure}

\begin{figure}
        \centering
        \subfloat[$d = 50\ \mu m$\label{fig:d502}]{\includegraphics[width=0.48\columnwidth]{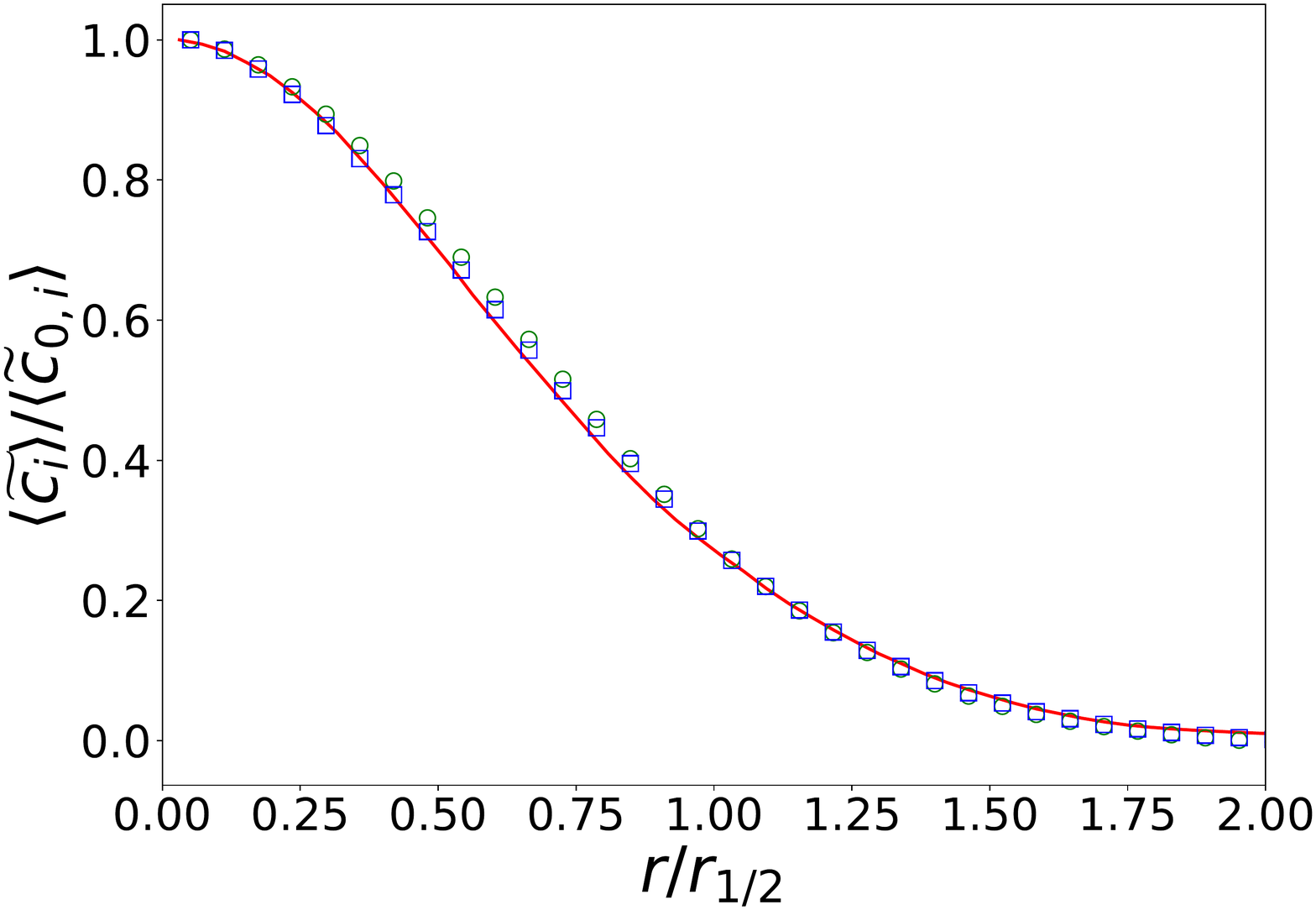}}
        \hfill
        \subfloat[$d = 360\ \mu m$ \label{fig:d3602}]{\includegraphics[width=0.48\columnwidth]{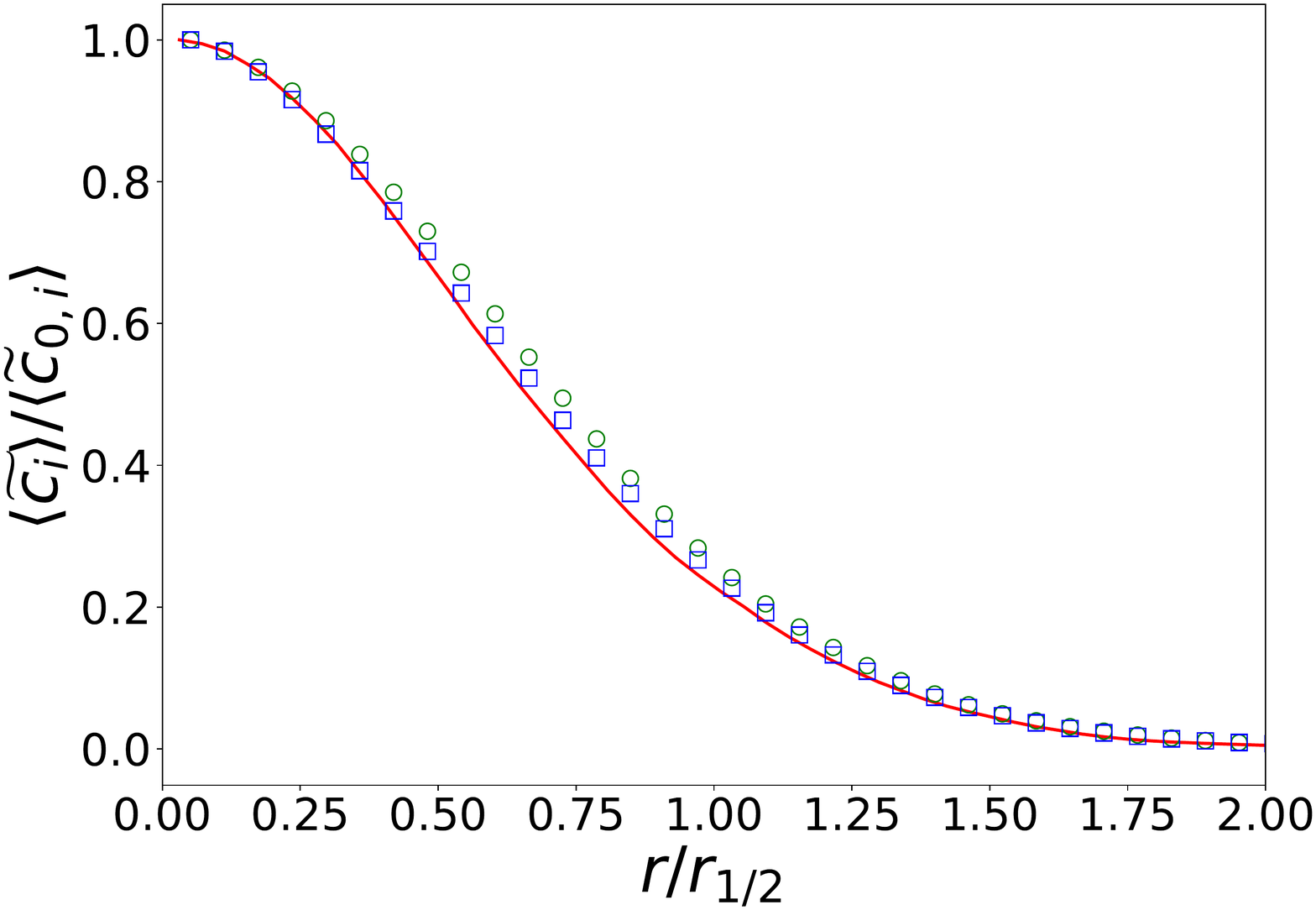}}\\
        \subfloat[$d = 683\ \mu m$ \label{fig:d6832}]{\includegraphics[width=0.48\columnwidth]{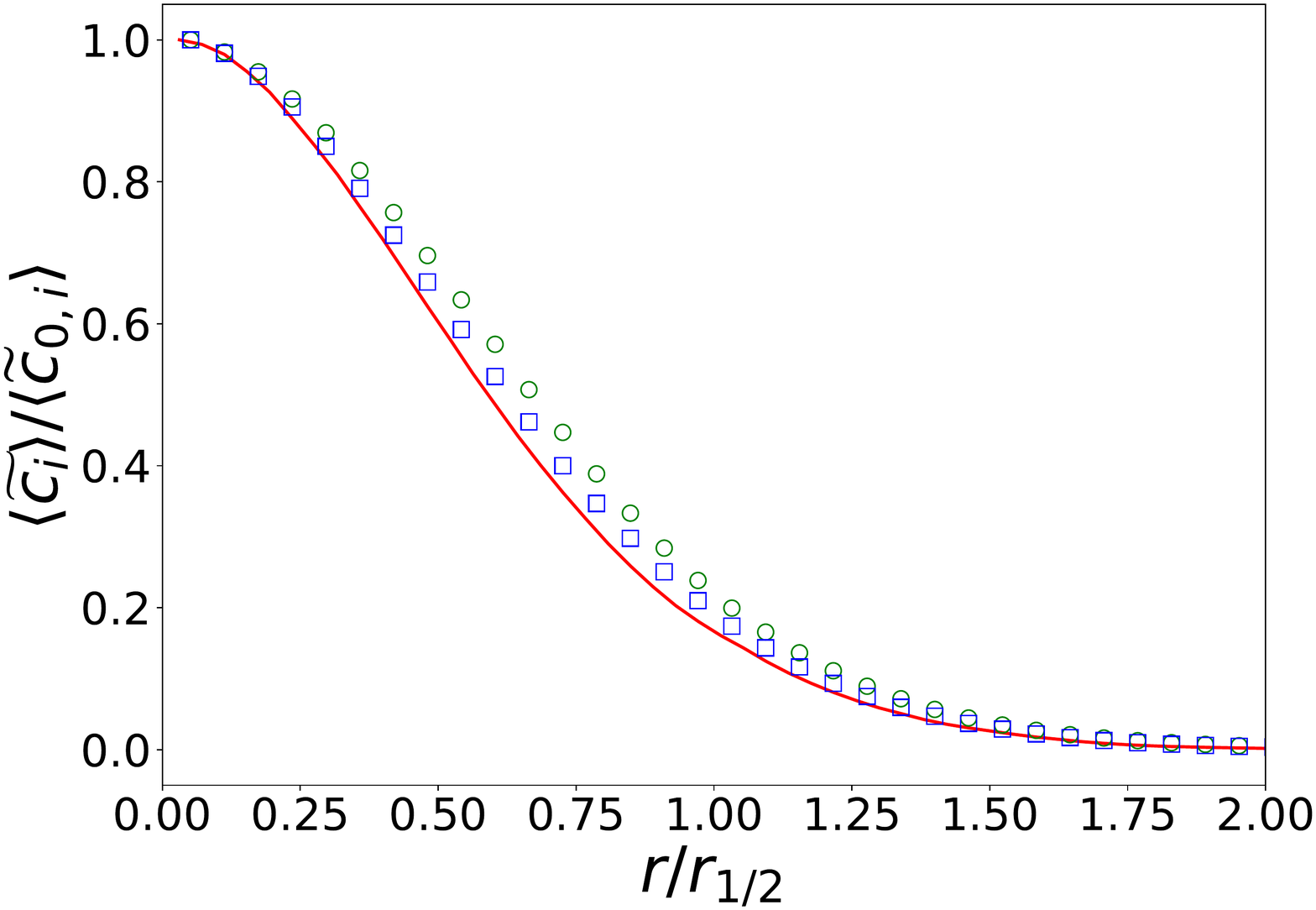}}
          \hfill
        \subfloat[$d = 1000\ \mu m$ \label{fig:d10002}]{\includegraphics[width=0.48\columnwidth]{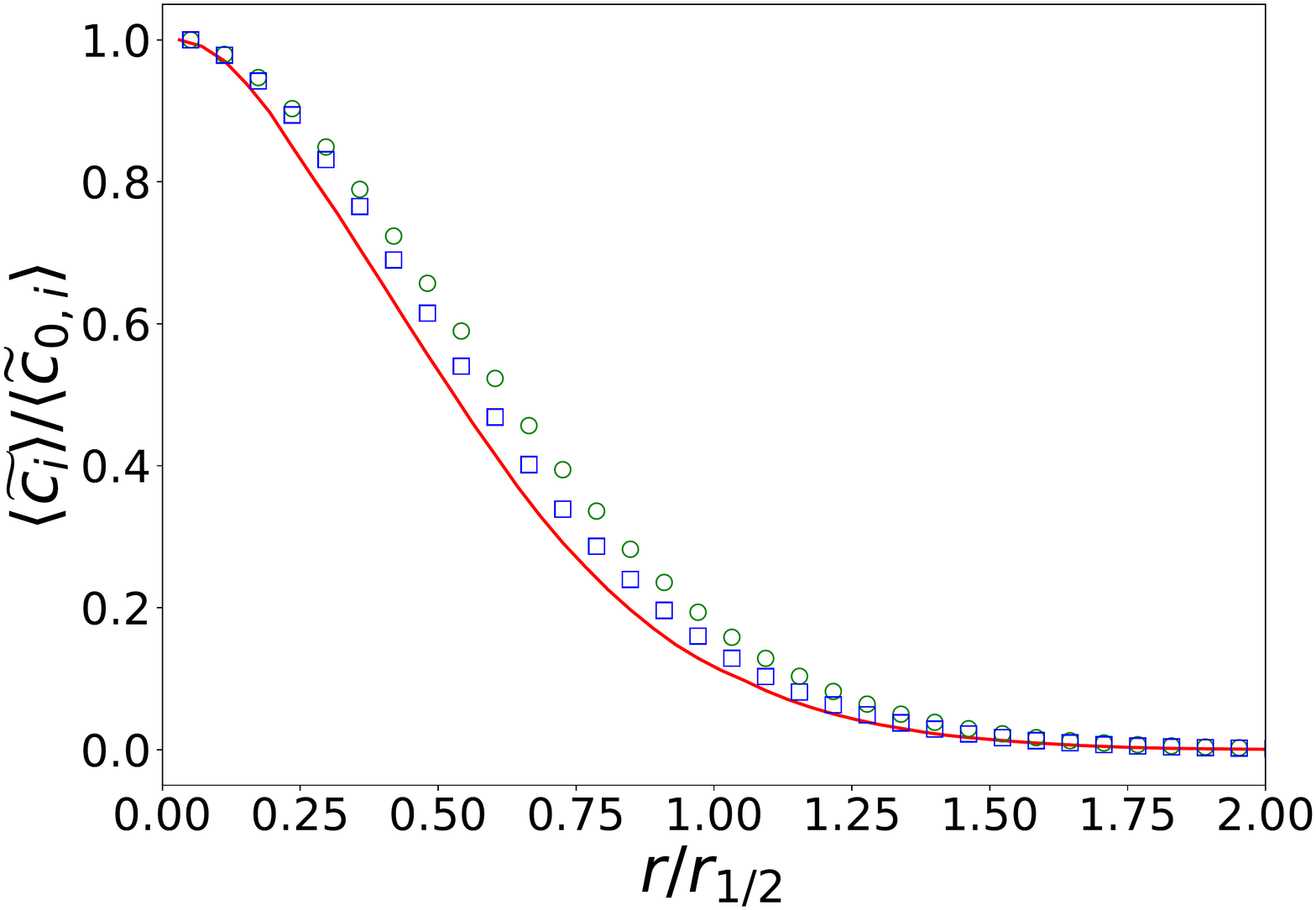}}
        
  \caption{Same as figure \ref{fig:conc_drop_sgs} at $z_{s}/D_J = 117$.}
    \label{fig:conc_drop_sgs2}
\end{figure}

We plot the radial concentration flux from CS1 in Figure \ref{fig:conc_flux_tot_coarse}  for two different downstream locations  $z_s/D_J = 80$ and $z_s/D_J = 115$, where $z_s = z - I_2$. We can see that for the coarse simulation, the subgrid flux closer to the injection location is $20\%$ of the total resolved flux. This suggests that the subgrid parameterization used for the concentration flux would be important. {Additionally, we show the subgrid component of the flux from FS1 and CS2. As expected, due to the higher resolution, the subgrid flux for FS1 is 4 times smaller than that of CS1. The subgrid flux for CS2, where the modified Schmidt number model (Equation (4.1) has been applied is also smaller due to reduction of the subgrid flux for larger droplet sizes.} We show the radial concentration distributions for the four droplet sizes at different downstream locations in Figures \ref{fig:conc_drop_sgs} and \ref{fig:conc_drop_sgs2}. Results show that the profiles for the smallest droplet sizes for all the simulations show good agreement and are quite independent of grid resolution. These droplets have a ratio of rise velocity to turbulent intensity  $W_{r,d_1}/w^{\prime} < O(0.01)$. For such low rise velocities there is negligible influence of the crossing trajectory effect. The difference is most evident for the larger droplet sizes, where due to the finite rise velocity, the particle diffusion is suppressed. We see that the modified Schmidt number model accurately corrects the distribution for the larger droplet size for the coarse simulation, to match with the higher resolution case by enforcing a higher Schmidt number for the subgrid flux calculation in LES. {The sensitivity of the simple choice of $C=1$ in Eq. (\ref{eqn:Sc_sgs}) in the LES is tested by calculating the effect on the modified Schmidt number at some representative position in the flow. At $z_s/D_J = 80$, we have $S_v = 0.7$. The differences in Schmidt number calculated for $C=0.5$ and $C=1.5$ at this location compared to the $C=1$ case used here are smaller than $15\%$. As can be seen, the sensitivity is not negligible but considering that the SGS flux is only a fraction of the total, the sensitivity of mean concentration profiles to this parameter is negligible, at least in the current study.}

\begin{figure}
    \centering
    \includegraphics[width = 0.8\columnwidth]{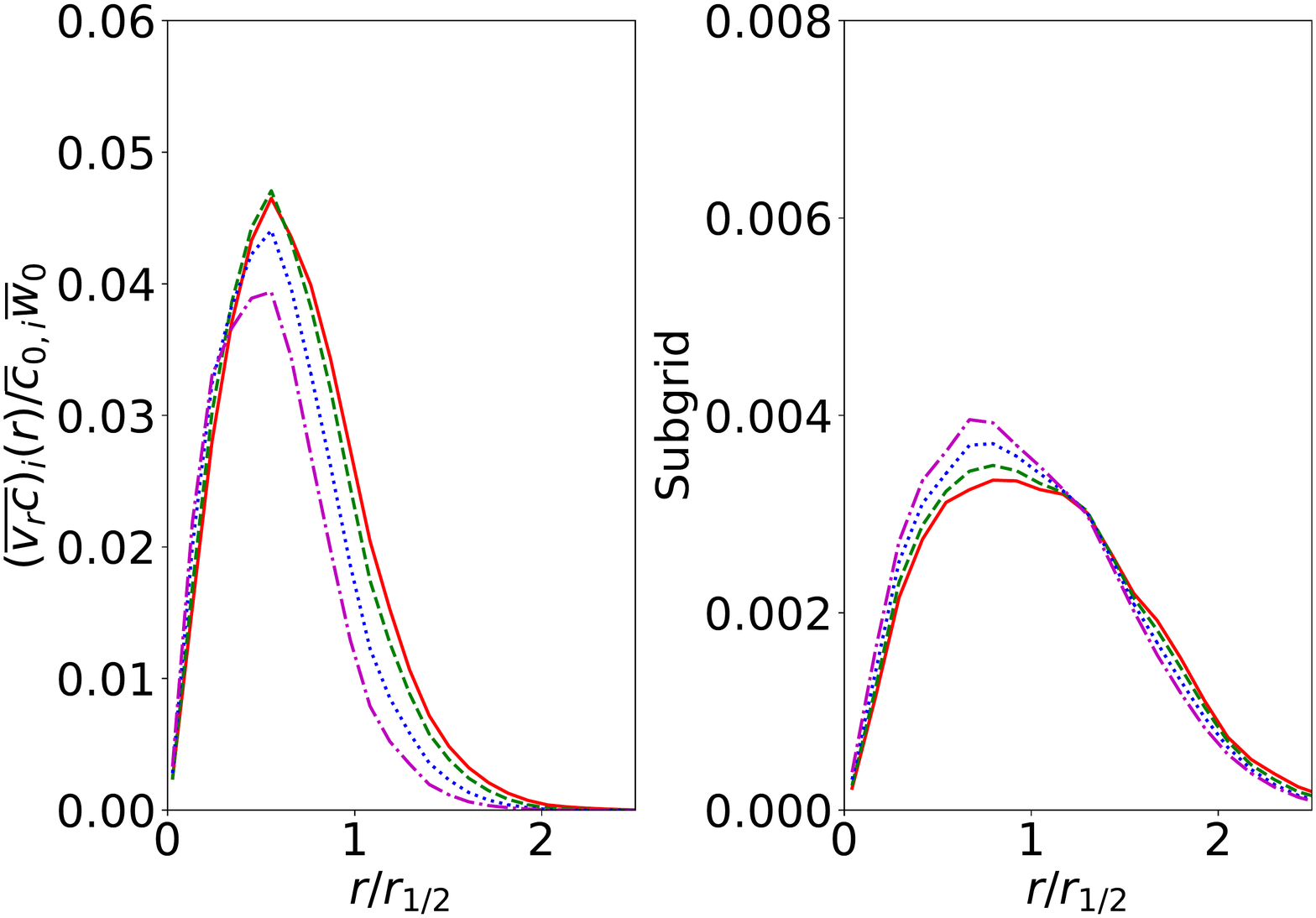}
       \caption{Normalized radial turbulent concentration flux and subgrid flux contribution for FS1 at a) $z_s/D_J = 140$  as a function of self similarity variable $r/r_{1/2}$. The lines are $d = 50\;\mu m$ (\protect\redline) $d = 366\ \mu m$ (\protect\greendline), $d = 683\ \mu m$ (\protect\blueddline) and $d = 1000\ \mu m$ (\protect\magdline).  }
    \label{fig:conc_flux_fine}
\end{figure}

\begin{figure}
    \centering
    \includegraphics[width = 0.8\columnwidth]{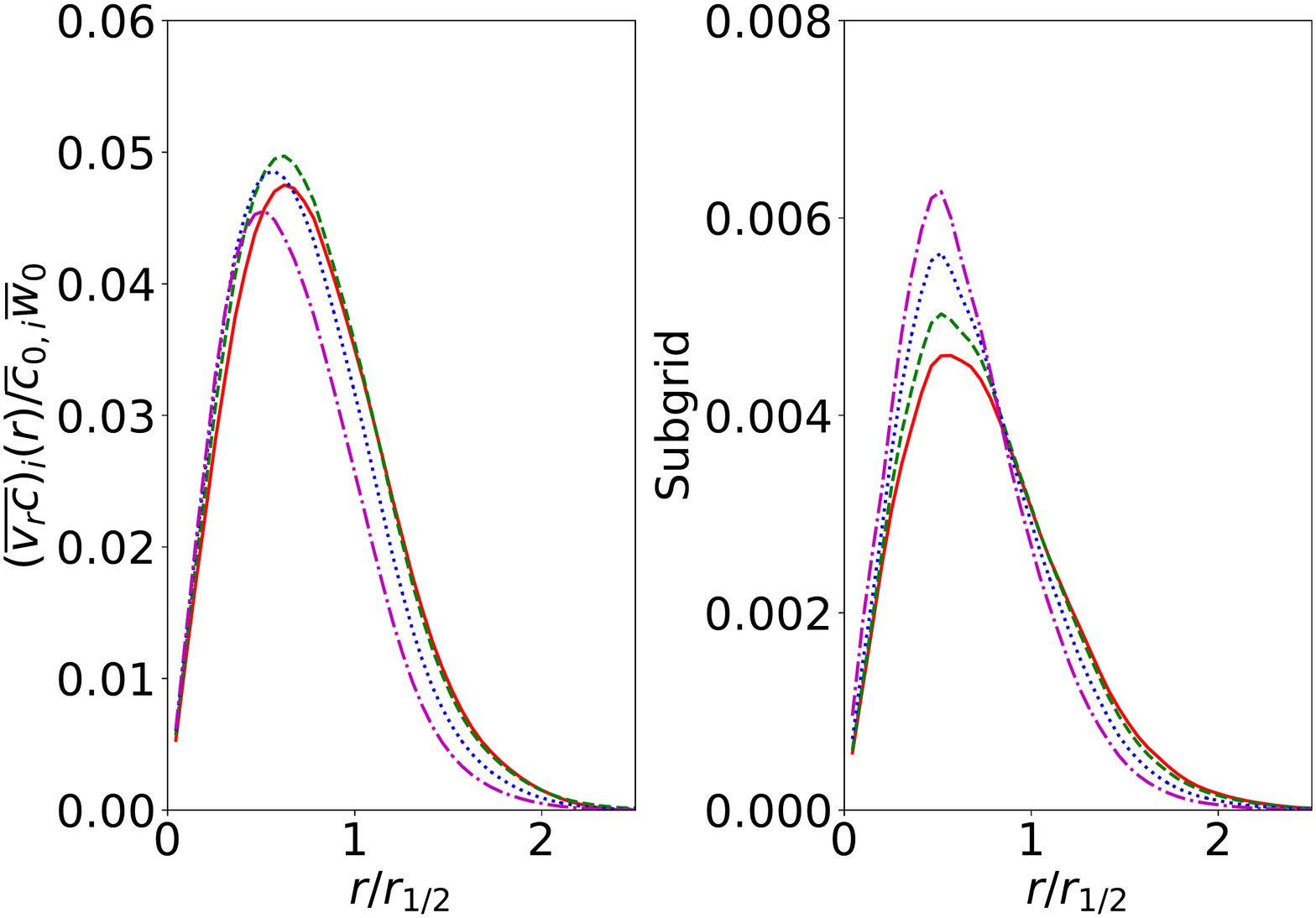}
    \caption{Normalized radial turbulent concentration flux and subgrid flux contribution for CS1 at a) $z_s/D_J = 140$  as a function of self similarity variable $r/r_{1/2}$. The lines are $d = 50\;\mu m$ (\protect\redline) $d = 366\ \mu m$ (\protect\greendline), $d = 683\ \mu m$ (\protect\blueddline) and $d = 1000\ \mu m$ (\protect\magdline). }
    \label{fig:conc_flux_c}
\end{figure}

\begin{figure}
    \centering
    \includegraphics[width = 0.8\columnwidth]{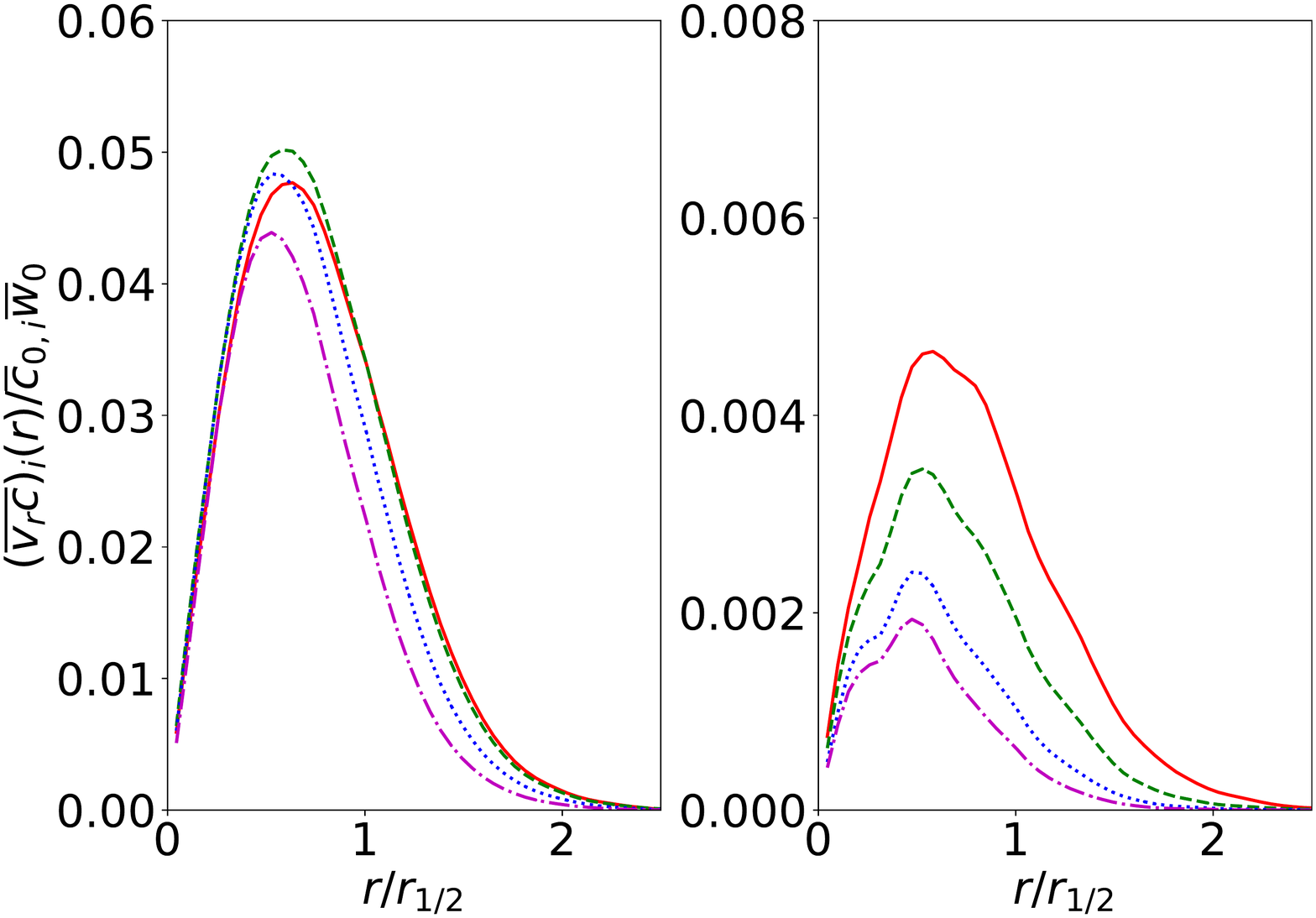}
      \caption{Normalized radial turbulent concentration flux and subgrid flux contribution for CS2 at  a) $z_s/D_J = 140$ as a function of self similarity variable $r/r_{1/2}$. The lines are $d = 50\;\mu m$ (\protect\redline) $d = 366\ \mu m$ (\protect\greendline), $d = 683\ \mu m$ (\protect\blueddline) and $d = 1000\ \mu m$ (\protect\magdline). }
    \label{fig:conc_flux_cc}
\end{figure}

In order to verify that the trends seen in mean concentration profiles are due to the differences in SGS modeled fluxes, we plot the turbulent concentration flux for the different droplet sizes in Figures \ref{fig:conc_flux_fine}, \ref{fig:conc_flux_c} and \ref{fig:conc_flux_cc}. The left panel depicts the resolved portion of the flux and the contribution due to the subgrid model is shown on the right panel. We can see that the subgrid flux is  smaller for the fine resolution case  compared to the coarser resolutions. The resolved flux $\langle \tilde{v}_r^{\prime}\tilde{c}^{\prime}\rangle$ follows a clear size-dependent trend for every grid resolution, with the flux being highest for the smallest drop and lowest for the largest drop. We observe a size dependence in the SGS flux with the contribution from the largest droplet being the highest. Simulations FS1 and CS1 use a constant Schmidt number SGS model for the concentration ensuring that the eddy diffusivity is relatively independent of droplet size. The observed minor differences in the droplet subgrid flux are due solely  to differences in the radial concentration gradient $\partial \tilde{c}_i/\partial r$ for the droplet sizes. The radial gradient of the concentration is inversely proportional to the profile half-width. As we have seen in section \S \ref{sec:Results}, the widths for the large droplets are reduced due to the crossing trajectory effect, resulting in a slightly higher gradient. Figure \ref{fig:conc_flux_cc} for CS2 shows that the subgrid flux contribution for the larger droplets is (correctly) suppressed as a higher Schmidt number is imposed by the new subgrid-scale  model. 

\section{Discussion and conclusions}\label{sec:Concl}

  We have explored the size based dispersion characteristics of buoyant droplets in a turbulent round jet using large eddy simulations. The droplet concentration field is modeled using the Equilibrium-Eulerian approach, where the droplet velocity is expanded as a function of the droplet time-scale, valid for small Stokes numbers. Droplets of different sizes (and thereby different rise velocities) are injected at the centerline of a turbulent round jet. The radial distributions of the droplet concentration fields were shown to be size dependent, with the width of the larger droplet sizes being suppressed due to crossing trajectory effects, when the more rapidly slipping particles cross resolved turbulent eddies, thus decreasing their interaction times with the eddies and reducing the net turbulent dispersion. The effect is smaller close to the jet injection location and becomes more pronounced further downstream. The Equilibrium-Eulerian approach with a constant SGS Schmidt number is effective in capturing the crossing trajectory effect when the grid resolution is sufficiently high and the subgrid contribution to the total flux is small.

For LES with coarser resolutions, where subgrid effects become noticeable, we proposed a  modified-Schmidt number SGS model based on the dispersion coefficient derived by \citet{Csanady1963}.
First, the \citet{Csanady1963} model was used to predict the evolution of the mean concentration half-width of different droplet sizes and comparisons with the  fine-resolution LES and showed good agreement. The modified-Schmidt number is defined as a function of the droplet rise velocity and the turbulence fluctuating velocity modelled based on the subgrid kinetic energy. The modified-Schmidt number SGS model is tested in a coarse LES of a turbulent round jet with droplets injected in the self-similar region downstream of the nozzle. The SGS model suppresses the transport of the larger droplet sizes by imposing a larger effective Schmidt number thereby reducing its subgrid contribution. 
Further improvements of the proposed SGS model are possible, such as dynamic versions based on test-filtering \citep{Germano91} to determine parameters that were fixed to unity in the present applications. 
{The proposed subgrid-scale diffusion model has been 
formulated so that it can be applied quite generally to any flow (as long as the particle Stokes number is small), i.e. Eq. (\ref{eqn:Sc_sgs}) does not depend on specific directions and it uses quantities typically available in LES. Nonetheless, since it has only been tested for the case of a round jet in this paper, we defer from making claims about general applicability until further applications to other flows can be made.}

\section*{Acknowledgements} 
Computational resources were provided by the Maryland Advanced Research Computing Center (MARCC). We also acknowledge XSEDE computing resources under grant \# ATM130032. A.A acknowledges Prof. Michael Mueller for support during the final stages of this work.

\section*{Declaration of Interests} The authors report no conflict of interest

\bibliographystyle{jfm}
\bibliography{references,references_2}

\end{document}